\documentclass[11pt,reqno]{amsart}  

\usepackage{euler}
\usepackage{amscd} 
\usepackage{mathrsfs}
\usepackage{amsmath,amsbsy,amsthm,amsfonts,amssymb,esint}
\usepackage[unicode,psdextra]{hyperref}
\usepackage{graphicx}
\usepackage{tikz}
\usepackage{color}

\newcommand{\DIFF}[1]{{\color{red} \bf #1}}

\usepackage[sort&compress,capitalise,nameinlink]{cleveref}
\crefname{section}{\textsection}{\textsection}
\crefname{subsection}{\textsection}{\textsection}
\usepackage{dsfont}
\usepackage{enumerate}

\newtheorem{theorem}{Theorem}[section]

\newtheorem{proposition}[theorem]{Proposition}
\newtheorem{corollary}[theorem]{Corollary} 
\theoremstyle{definition}

\newtheorem{remark}[theorem]{Remark} 
\numberwithin{equation}{section}

\newcommand{\one}{\mathds{1}}

\newcommand{\xv}{\mathbf{x}}

\newcommand{\uv}{\mathbf{u}}
\newcommand{\vv}{\mathbf{v}}

\newcommand{\dom}{\mathscr{D}}
\newcommand{\diff}{\mathrm{d}}

\newcommand{\la}{\lambda}

\newcommand{\supp}{\mathrm{supp}}
\newcommand{\avg}[1]{\lf\langle #1 \ri\rangle}

\newcommand{\beq}{\begin{equation}}
\newcommand{\eeq}{\end{equation}}

\newcommand{\OO}{\mathcal{O}}

\newcommand{\Z}{\mathbb{Z}}

\newcommand{\R}{\mathbb{R}}

\newcommand{\bdm}{\begin{displaymath}}
\newcommand{\edm}{\end{displaymath}}
\newcommand{\bdn}{\begin{eqnarray}}
\newcommand{\edn}{\end{eqnarray}}
\newcommand{\bay}{\begin{array}{c}}
\newcommand{\eay}{\end{array}}
\newcommand{\ben}{\begin{enumerate}}
\newcommand{\een}{\end{enumerate}}
\newcommand{\beqn}{\begin{eqnarray}}
\newcommand{\eeqn}{\end{eqnarray}}
\newcommand{\bml}[1]{\begin{multline} #1 \end{multline}}
\newcommand{\bmln}[1]{\begin{multline*} #1 \end{multline*}}

\newcommand{\lf}{\left}
\newcommand{\ri}{\right}
\newcommand{\disp}{\displaystyle}
\newcommand{\tx}{\textstyle}

\newcommand{\bra}[1]{\lf\langle #1\ri|}
\newcommand{\ket}[1]{\lf|#1 \ri\rangle}
\newcommand{\braket}[2]{\lf\langle #1|#2 \ri\rangle}
\newcommand{\braketr}[2]{\lf\langle #1\lf|#2\ri. \ri\rangle}
\newcommand{\braketl}[2]{\lf.\lf\langle #1\ri|#2 \ri\rangle}
\newcommand{\mean}[3]{\bra{#1}#2\ket{#3}}
\newcommand{\meanlrlr}[3]{\lf\langle #1\lf|#2\ri|#3\ri\rangle}

\renewcommand{\leq}{\leqslant}
\renewcommand{\geq}{\geqslant}

\title[Magnetic Perturbations of Anyonic and  Aharonov-Bohm Schr\"{o}dinger Operators]{Magnetic Perturbations of Anyonic and  Aharonov-Bohm Schr\"{o}dinger Operators}

\author{Michele Correggi}
\address{Dipartimento di Matematica, Politecnico di Milano, P.zza Leonardo da Vinci, 32, 20133, Milano, Italy}
\email{\href{mailto:michele.correggi@gmail.com}{michele.correggi@gmail.com}}
\urladdr{\href{https://sites.google.com/view/michele-correggi}{https://sites.google.com/view/michele-correggi}}

\author{Davide Fermi}
\address{Scuola Normale Superiore, Piazza dei Cavalieri, 7, I-56126, Pisa, Italy}
\email{\href{mailto:fermidavide@gmail.it}{fermidavide@gmail.it}}
\urladdr{\href{https://fermidavide.com}{https://fermidavide.com}}

\begin{document}

\begin{abstract} 
	We study the Hamiltonian describing two anyons moving in a plane in presence of an external magnetic field and identify a one-parameter family of self-adjoint realizations of the corresponding Schr\"{o}dinger operator. We also discuss the associated model describing a quantum particle immersed in a magnetic field with a local Aharonov-Bohm singularity. For a special class of magnetic potentials, we provide a complete classification of all possible self-adjoint extensions. 
\end{abstract}


\maketitle

\section{Introduction and Main Results}

The possibility of particles obeying exotic statistics  in lower dimensions -- named {\it anyons} by Wilczek in the `80s \cite{Wi} --  is known since the late `70s \cite{LM} (see also \cite{GMS1,GMS2}), although only later effective models with anyonic quasi-particles have been proposed to describe the fractional quantum Hall effect \cite{ASW}. Amid later investigations, let us mention in particular \cite{AHK,CLL,GHKL,M,T1,T2}, dealing with finite-size ``regularized'' anyons, and \cite{MHV,V}, studying 2-body anyonic systems with Coulomb interaction. It is noteworthy that many of these works also consider the presence of an external magnetic field or an harmonic trapping potential.
In the last years 
this line of research has been renewed with contributions about the emergence of anyonic behaviors in condensed matter physics \cite{BLLY,Eetal,LunRou2,Yetal}, as well as more mathematically oriented works \cite{CDLR,CLR,CoOd,Gi,AK,LLN,LQ,LunRou1,LSe,LSo} on the characteristic features of anyonic systems. 

In several of the above physical investigations the anyonic system is immersed in an external magnetic field or, more generally, the presence of a magnetic field is key for the emergence of the anyonic behavior. It is however remarkable that, with the exception of \cite{Gi} and some very special magnetic fields admitting an explicitly solution, all the rigorous results apply to ``free'' or at most trapped anyons, {\it i.e.}, without any interaction and in absence of applied magnetic fields. In this paper we deal precisely with the latter question and study the self-adjoint realizations of the Schr\"{o}dinger operator for two non-interacting\footnote{Here, we mean by ``non-interacting anyons'' that there are no interactions besides those induced by the anyonic statistics, although the magnetic field itself may be viewed as the result of a magnetic dipole coupling between the particles.} anyons in a magnetic field. It is indeed well known that there are infinitely many ways of implementing the dynamics of two-anyon systems \cite{BDG,BoS,CoOd,DR,Ou}, since the naive symmetric operator providing the energy admits a one-parameter family of self-adjoint extensions. This can be seen by directly addressing the quadratic form associated with the formal operator \cite{CoOd} or studying the effective model on the half-line after the decomposition in spherical harmonics \cite{BDG,DR} (see also \cite{DR18,DR20} and the discussion below). Whether such a behavior is affected by the presence of the magnetic field is the main question we answer in this work. Note indeed that, even in \cite{Gi}, where the effects of the magnetic field are taken into account in the study of a suitable almost-bosonic large-$ N $ limit of a many-anyon system, only the Friedrichs realization of the Schr\"{o}dinger operator is considered.

Another important motivation to investigate one-particle anyonic Schr\"{o}dinger operators in presence of an external magnetic field is related to the perspective of implementing many-anyon Hamiltonians as suitable self-adjoint realizations. An intermediate step towards this goal, which we plan to discuss in a future paper, is indeed the analysis of the Schr\"{o}dinger operator of a particle moving on a plane with finitely many Aharonov-Bohm fluxes: the connection with the present investigation relies on the fact that, in a neighborhood of the center of a given flux, the magnetic potential associated with the other fluxes is regular and therefore one locally recovers an operator of the form \eqref{HalS} below. There is however an important key difference with the following discussion, since it is impossible to require that such a magnetic potential vanishes at the point where the singularity is located, as we assume in next \eqref{S0}. We refer to the forthcoming work for further details on this question.

Before introducing the details of the mathematical setting, we remark that the model describing two non-interacting anyons is strongly connected with the physics of a quantum particle moving in a Aharonov-Bohm magnetic field \cite{AdTe,Dab}, {\it i.e.}, in presence of a magnetic potential with a typical $ 1/|\xv| $ singularity at the origin. Formally, when the anyonic energy is written in the magnetic gauge and up to the extraction of the center of mass motion (see, {\it e.g.}, \cite[Sect. 1]{CoOd}), the two Schr\"{o}dinger operators coincide up to Kato-small corrections, but an important difference is due to the symmetry constraint on anyonic wave functions, which must actually behave as bosonic functions in the magnetic gauge. Concretely, if $ \xv \in \R^2 $ stands for the relative distance of two anyons or the position of the Aharonov-Bohm particle, the formal Schr\"{o}dinger operator reads in both cases
\begin{equation}\label{Hal0}
	H_{\alpha,0} := \big(- i \nabla + \mathbf{A}_{\alpha})^{2}, \qquad \mathbf{A}_{\alpha}(\mathbf{x}) := \alpha\;{\mathbf{x}^{\perp} \over |\mathbf{x}|^2}\,,
\end{equation}
where we have used the notation $\mathbf{x}^{\perp} : = (-y,x)$, for $\mathbf{x} = (x,y) \!\in\! \mathbb{R}^2$, and 
\beq 
	\label{eq: alpha}
	\alpha \in (0,1) 
\eeq
 is the statistic parameter in the anyonic case (so that $ \alpha = 0 $ and $ \alpha = 1 $ identify bosons and fermions, respectively) or a measure of the magnetic flux otherwise. The difference between the two models is apparent in the space of states, which is the whole of  $ L^2(\R^2)$ for the Aharonov-Bohm particle, while it must be restricted to $ L_{\mathrm{even}}^2(\R^2) $ for anyons, where $ L^2_{\mathrm{even}}(\R^2) : = \lf\{ \psi \in L^2(\R^2) \: | \: \psi(-\xv) = \psi(\xv) \ri\} $. Note that the restriction to a real parameter $ \alpha $ varying in $ (0,1) $ is motivated by periodicity over the angular variable $ \vartheta $, which can be made apparent by a decomposition in cylindrical harmonics (see \eqref{RepSph}). In fact, in the Aharonov-Bohm case, one could exploit another symmetry of the model, {\it i.e.}, the invariance under complex conjugation, to further restrict $ \alpha \in (0,1/2] $, but such a symmetry is obviously broken by the presence of an external magnetic field. Therefore, we will stick to the choice \eqref{eq: alpha}.
 
We also remark that, after the decomposition in cylindrical harmonics, i.e., setting
\bdm
	L^2(\R^2) = \disp\bigoplus_{k \in \Z_\mathrm{even}} \mathscr{H}_k, 	\qquad		\mathscr{H}_k : = L^2(\R^+) \otimes \mathrm{span} \lf\{ \tx\frac{1}{\sqrt{2\pi}} e^{i k \vartheta} \ri\},
\edm
the operator $ H_{\alpha,0} $ becomes diagonal and, in each subspace with given angular momentum $ k \in \Z $, one finds an effective operator acting on $ L^2(\R^+) $, which is unitarily equivalent to
\beq
	L_{a_k} : = - \partial_r^2 + \frac{a_k^2}{r^2},
\eeq
where $ a_k = \alpha + k $. It turns out \cite{BDG,DR} that the self-adjoint extensions of such an operator can be fully classified and basically all of its spectral and scattering properties explicitly derived: a very convenient approach starts with the identification of the maximal and minimal extensions, which allows to construct all the other extensions by restricting the maximal operator via a suitable boundary condition at the origin. This approach is certainly very efficient but does not extend straightforwardly in presence of an interaction and/or a trapping potential, which are not rotationally invariant or which are not regular enough. To include such cases, one can take a different view point and consider instead the quadratic form $ \mean{\Psi}{H_{\alpha}}{\Psi} $, as done in \cite{CoOd} and below.

In this paper we plan to investigate the perturbation of the operator \eqref{Hal0} induced by an external magnetic field $ \mathbf{B}(\xv) $ perpendicular to the plane where the particles move. Since we aim at covering  discontinuous magnetic fields, {\it e.g.}, magnetic barriers or magnetic step potentials, which are known to play an interesting role due to the occurrence of edge currents, we assume that $ \mathbf{B} $ is obtained as the (weak) rotation of a magnetic potential $ \mathbf{S}: \mathbb{R}^2 \to \mathbb{R}^2  $, {\it i.e.}, $ \mathbf{B}(\xv) = \lf( \nabla \times \mathbf{S} \ri) (\xv)  $, which has a minimal amount of regularity for the model to be meaningful. More precisely, we consider the Hamiltonian operators determined by various self-adjoint realizations in $L^2(\mathbb{R}^2)$ of the differential operator
\begin{equation}\label{HalS}
H_{\alpha,S} := (- i \nabla + \mathbf{A}_{\alpha} + \mathbf{S})^{2} \,,
\end{equation}
with $\mathbf{S}$ locally bounded on $\mathbb{R}^2$: 
\begin{equation}
	\mathbf{S} \in L^{\infty}_{\mathrm{loc}}(\mathbb{R}^2)\,. 
	\label{Shyp}
\end{equation}

If $ \mathbf{S} $ has a very specific form, i.e., $ \mathbf{S} = {S}_{\perp} \hat{\xv}^{\perp} $, for some constant $ {S}_{\perp} \in \R$, the operator $ H_{\alpha,S} $ admits a diagonalization similar to the case $ \mathbf{S} = 0 $ and the corresponding effective operator on the half-line was studied in full generality in \cite{DR18,DR20}. Here however we mostly aim at considering more general magnetic potentials (in particular not rotationally invariant and possibly less regular), which prevents us to adopt the boundary condition approach followed in \cite{BDG,DR,DR18,DR20}. In \cref{sec: special}, however, we are going to show that if one imposes very restrictive conditions on $ \mathbf{S} $ (yet more general than those mentioned above), then it is possible to extract much more information.

We indeed anticipate that we are not able to solve the self-adjoint extension problem in the full generality of the above assumption \eqref{Shyp}. We indeed present and prove the following results:
\begin{itemize}
	\item if $ \mathbf{S} $ is regular enough at the origin, {\it i.e.},
	\bdm
	\mathbf{S}\in
	\begin{cases}
		C^{0}(B_{r_{0}})\,, 	&	\mbox{for } \alpha \in (0,1/2)\,,		\\
		C^{0,\nu}(B_{r_0})\,, \mbox{ for some } \nu  > 2\alpha -1\,, & \mbox{for } \alpha \in [1/2,1)\,,
	\end{cases}
	\edm
	we explicitly construct (\cref{sec: regular}) a one-parameter family of self-adjoint operators realizing \eqref{HalS}, which, if in addition $ \mathbf{S} \in L^{\infty}(\R^2) $, exhaust all possible self-adjoint realizations of $H_{\alpha,S}$ in $ L^2_{\mathrm{even}}(\R^2) $;
	\item the same result is proven also (\cref{sec: discontinuous}) under weaker conditions on $ \mathbf{S} $, {\it i.e.}, assuming that  the projections $ S_{\parallel} : = \mathbf{S} \cdot \hat{\mathbf{x}}$ and $ S_{\perp} : = \mathbf{S} \cdot \hat{\mathbf{x}}^{\perp}$ are continuous at the origin:
	\bdm
	S_{\perp}, S_{\parallel} \in
	\begin{cases}
		C^{0}(B_{r_{0}})\,, 	&	\mbox{for } \alpha \in (0,1/2)\,,		\\
		C^{0,\nu}(B_{r_0})\,, \mbox{ for some } \nu > \alpha \,, & \mbox{for } \alpha \in [1/2,1)\,,
	\end{cases}
	\qquad
	\lim_{\xv \to 0} S_{\parallel}(\xv) = 0;
	\edm
	\item finally, if $ \mathbf{S} $ has a very special form, {\it i.e.}, $ \mathbf{S}(\mathbf{x}) = S\lf(|\mathbf{x}|\ri)\, \hat{\mathbf{x}}^{\perp} $, for some $ S \in C^{0,1}\big([0,r_0)\big) \cap L^{\infty}(\mathbb{R}_{+}) $, we show  (\cref{sec: special}) that the operator \eqref{HalS} can be reduced to a sum of perturbations of the {\it Whittaker operators} studied in \cite{DR,DR18,DR20} and we provide a complete classification of all self-adjoint Schr\"{o}dinger operators associated with a particle moving on a plane with a Aharonov-Bohm flux at the origin.
\end{itemize}

It is worth to remark that, in the generality of the first case, we have no access to the defect functions of the perturbed operator and therefore, at least for continuous potentials $ \mathbf{S} $, our constructions involve the defect function of the unperturbed Aharonov-Bohm Hamiltonian with $ \mathbf{S} = \mathbf{0} $ (see \eqref{greenexp}), multiplied by a smooth cut-off. However, if $ \mathbf{S} \cdot \hat{\mathbf{x}}^{\perp}$ is continuous at $ \mathbf{0} $ and does not vanish there (which makes $ \mathbf{S} $ discontinuous at the origin), such a function does not have the correct asymptotic behavior at the origin and a suitable modification of its sub-leading order term in the asymptotics is called for.

\subsection{The Friedrichs extension and the $ \mathbf{S} = \mathbf{0} $ case}

As usual, before dealing with the self-adjoint realizations of $ H_{\alpha,S} $, it is key to identify the Friedrichs extension and possibly characterize its properties. We thus introduce the quadratic form associated with $ H_{\alpha,S} $ as
\begin{equation}\label{QalS}
Q_{\alpha,S}[\psi] := \!\int_{\mathbb{R}^2}\! d\mathbf{x}\; \lf| \lf(- i \nabla +  \mathbf{A}_{\alpha} +  \mathbf{S} \ri) \psi \ri|^{2} \,,
\end{equation}
which is well defined at least on smooth functions with compact support away from the origin. For later reference, we also denote by $\dot{H}_{\alpha,S}$ the closure of the symmetric operator $H_{\alpha,S} \!\upharpoonright\! C^{\infty}_{c}(\mathbb{R}^2 \setminus \{\mathbf{0}\})$, which is densely defined on $L^2(\mathbb{R}^2)$. Its domain is given by
\begin{equation}
		\dom\big(\dot{H}_{\alpha,S}\big) = \lf\{\psi \in L^2(\mathbb{R}^2) \,\Big|\,\psi \in H^2_{\mathrm{loc}}(\mathbb{R}^2\setminus \{\mathbf{0}\})\,,\;H_{\alpha,S}\, \psi \in L^2(\mathbb{R}^2) \ri\}\,. \label{Hdot}
\end{equation} 
Next, we observe that $\|\psi\|_{\alpha,S}^2 := \|\psi\|_{2}^{2} + Q_{\alpha,S}[\psi]$ defines a norm and therefore we can identify the quadratic form associated with the Friedrichs extension of $ \dot{H}_{\alpha,S} $ with
\begin{equation}\label{QFr}
	Q_{\alpha,S}^{(F)}[\psi] := Q_{\alpha,S}[\psi], \qquad \mbox{for all } \psi \!\in\! \dom \big[Q_{\alpha,S}^{(F)} \big] := \overline{C^{\infty}_{\mathrm{c}}(\mathbb{R}^2\!\setminus\! \{\mathbf{0}\})}^{\;\|\cdot\|_{\alpha,S}}\,.
\end{equation}

	\begin{proposition}[Friedrichs extension]	
		\label{prop:QF}
		\mbox{}	\\ 
		Let $\alpha \in (0,1)$ and $ \mathbf{S} \in L^{\infty}_{\mathrm{loc}}(\mathbb{R}^2)$. Then:
		\begin{enumerate}[i)]
			\item The quadratic form $Q_{\alpha,S}^{(F)}$ is closed and non-negative on its domain and
				\begin{equation}\label{domQF}
					\dom \big[Q_{\alpha,S}^{(F)}\big] = \lf\{\psi \in L^2(\mathbb{R}^2)\; \big|\;(-i\nabla + \mathbf{S})\, \psi \in L^2(\mathbb{R}^2)\,,\; \mathbf{A}_{\alpha} \psi \in L^2(\mathbb{R}^2) \ri\} \,.
				\end{equation}
				Moreover, for any $\psi \in \dom\big[Q_{\alpha,S}^{(F)}\big]$ there holds\footnote{Here, $B_{r} \equiv B_{r}(\mathbf{0}) : = \lf\{\mathbf{x} \in \mathbb{R}^2\,|\;|\mathbf{x}| < r \ri\} $ stands for the open disc with center at the origin and radius $r > 0$ and we denote by $ \diff \Sigma_r : = r \diff \vartheta $ the measure on the circle $\partial B_r$ induced by the usual Lebesgue measure on $\mathbb{R}^2$.}
				\begin{equation}
					\lim_{r \to 0^+} \lf( r^{-1} \int_{\partial B_{r}}  d\Sigma_r\;|\psi|^2 \ri) = 0 \,, \qquad
			\lim_{r \to 0^+} \left( r \int_{\partial B_{r}}\hspace{-0.2cm} d\Sigma_r\;|\partial_{r}\psi|^2\right) = 0 \,. \label{limdomQF}
				\end{equation}
			\item The self-adjoint operator $H_{\alpha,S}^{(F)}$ associated with $Q_{\alpha,S}^{(F)}$ acts as $H_{\alpha,S}$ on its domain
				\begin{equation}
					\dom\big(H_{\alpha,S}^{(F)}\big) = \lf\{\psi \in \dom \big[ Q_{\alpha,S}^{(F)} \big] \;\Big|\; H_{\alpha,S}\,\psi \in L^2(\mathbb{R}^2) \ri\}\,. 							\label{domHF}
				\end{equation}
		\end{enumerate}
	\end{proposition}
	
	\begin{remark}[Bounded magnetic potentials]
		\label{rem: bounded S}
		\mbox{}	\\
		If $ \mathbf{S} $ is bounded also at infinity, it is not difficult to realize (see \cref{sec: friedrichs proof}) that \eqref{domQF} and \eqref{domHF} imply that $ \mathbf{S} $ does not affect the domain, {\it i.e.}, if $ \mathbf{S} \in L^{\infty}(\R^2) $, then
		\begin{equation}
			\dom \big[ Q_{\alpha,S}^{(F)} \big]
			 = \dom \big[ Q_{\alpha,0}^{(F)}\big]\,, \qquad
			\dom \big(H_{\alpha,S}^{(F)}\big) = \dom \big(H_{\alpha,0}^{(F)}\big)\,. \label{domHF0}
		\end{equation}
	\end{remark}
	
	\begin{remark}[Regularity of $ \mathbf{S} $]
		\label{rem:QF} 
		\mbox{}	\\
		The regularity of $\mathbf{S}$ at the origin plays no role in the derivation of the above results and, indeed, assuming that $S \!\in\! L^{\infty}_{\mathrm{loc}}(\mathbb{R}^2)$ is perfectly sufficient. In this case, whenever $\psi \!\in\! \dom (Q_{\alpha,S}^{(F)})$, one also gets, via \eqref{domQF}, that $\psi \!\in\! H^{1}_{\mathrm{loc}}(\mathbb{R}^2)$ and $(-i\nabla\!+\! \mathbf{A}_{\alpha})\,\psi \!\in\! L^{2}_{\mathrm{loc}}(\mathbb{R}^2)$.
\end{remark}

We now recall what happens in absence of additional magnetic fields, {\it i.e.}, for $ \mathbf{S} = \mathbf{0} $. There are in fact several ways to realize the formal expression \eqref{Hal0} as a self-adjoint operator. Following \cite{BDG,DR}, we expose here the most direct way of doing that (see also \cite{CoOd} and below for an alternative approach): one first considers the operator $ H_{\alpha,0} $ on the domain of smooth functions with support away from the origin and constructs its adjoint $ H_{\alpha,0}^* $, whose domain contains functions $ \psi $ which are $ H^2 $ outside the origin but may diverge there, in such a way that $ H_{\alpha,0} \psi \in L^2(\R^2) $, where the latter stands for the formal action of \eqref{Hal0}. More precisely, it turns out \cite[Sect. 2.2]{DR} that any $ \psi \in  \dom(H_{\alpha,0}^*) $ has an asymptotic behavior at the origin which may contain terms of order $ r^{\pm \alpha} $, on top of the usual $ H^2 $ regular terms. Note that, if this is the case, $ H_{\alpha,0} \psi \in L^2(\R^2) $ thanks to the cancellation of divergences, but $ \Delta \psi $ and $ \mathbf{A}_{\alpha} \psi $ do not belong to $ L^2(\R^2) $ separately. Next, one looks for the self-adjoint extensions of $ H_{\alpha,0} $ obtained by restricting the domain of the adjoint via the introduction of a suitable boundary condition at the origin, or, equivalently, by imposing a precise asymptotic behavior there. This leads to the one-parameter family $ \big\{ H_{\alpha,0}^{(\beta)} \big\}_{\beta \in \R \cup \lf\{ +\infty \ri\}} $ of self-adjoint extensions acting as $ H_{\alpha,0} $ on the domain
\bml{
	\dom\big( H_{\alpha,0}^{(\beta)} \big) =  \lf\{ \psi \in L^2_{\mathrm{even}}(\R^2) \: \Big| \: H_{\alpha,0}\, \psi \in L^2_{\mathrm{even}}(\R^2)\,, \;  \ri. 	\\
	\lf. \avg{\psi}(r) \underset{ r \to 0^{+}}{\sim} C \left(r^{\alpha} + {2^{2\alpha}\, \pi\, \alpha\,\Gamma^2(\alpha) \over \beta}\,r^{-\alpha} \right) + o(r), \;\mbox{for some } C \in \R \ri\},
\label{eq: dom hamb alt}
}
where we have denoted by $ \avg{ f}: \R^+ \to \mathbb{C} $ the angular average of a function $ f: \R^2 \to \mathbb{C} $ for short, {\it i.e.}, 
\beq
	\label{eq: avg}
	\avg{f}(r) : = \frac{1}{2\pi r} \int_{\partial B_r} \diff \Sigma_r \: f = \frac{1}{2 \pi} \int_0^{2\pi} \diff \vartheta \: f\big(\xv(r,\vartheta)\big)\,,
\eeq 
and chosen the parameter labelling the extension for later convenience\footnote{The parameters $ m \equiv m_{\mbox{{\tiny \cite{DR}}}}, \kappa \equiv \kappa_{\mbox{{\tiny \cite{DR}}}} $ in \cite[Eq. (2.3)]{DR} are given in terms of $\alpha,\beta$ as $
m_{\mbox{{\tiny \cite{DR}}}} = \alpha\, $, $ \kappa_{\mbox{{\tiny \cite{DR}}}}  = {2^{2\alpha}\, \pi\, \alpha\,\Gamma^2(\alpha) \over \beta}\, $.}. Note that, with such a choice, the case $ \beta = +\infty $ is included in the family and recovers the Friedrichs extension of \cref{prop:QF}, since the singular term $ r^{-\alpha} $ in the expansion at the origin disappears.
  
The same result can be obtained in a more elaborate way by considering the quadratic form associated with the operator $ H_{0,\alpha} $ and its extensions (see \cite{CoOd}). We expose this approach too, since it is more suited for our further investigation and we are going to follow it in the rest of the paper. The main advantage is indeed that such an approach is independent of the rotational symmetry of the model. However, it requires to introduce the defect function $G_{\lambda} \in L^2(\mathbb{R}^2)$, $\lambda > 0$, as the unique radial solution of the equation 
\begin{equation}
\lf((-i \nabla + \mathbf{A}_{\alpha})^2 + \lambda^2 \ri)\, G_{\lambda} = 0, \qquad \mbox{in\, $\mathbb{R}^2 \setminus \{\mathbf{0}\}$}\,, \label{greeeq}
\end{equation}
{\it i.e.}, the radial function
\begin{equation}
	G_{\lambda}(r) = \lambda^{\alpha}\, K_{\alpha}(\lambda\,r) \,, \label{greenexp}
\end{equation}
where $K_{\alpha}$ is the modified Bessel function of order $\alpha$. Let us remark that $G_{\lambda}$ is real-valued and square-integrable \cite[Eq.\,6.521.3]{Grad} and
\begin{equation}\label{G2norm}
\|G_{\lambda}\|_{2}^{2} = \frac{\pi^2 \alpha \lambda^{2\alpha-2}}{\sin(\pi\,\alpha)} = : \alpha\, c_{\alpha}\, \lambda^{2\alpha-2} \,.
\end{equation}
Furthermore, its asymptotic expansions are given by \cite[\S 10.31 and Eq.\,10.40.2]{NIST}
\begin{gather}
G_{\lambda}(r) = \left\{\!\begin{array}{ll}
\displaystyle{{\Gamma(\alpha) \over 2^{1 - \alpha}}\, r^{-\alpha} - {\Gamma(1-\alpha)\,\lambda^{2 \alpha} \over 2^{1 + \alpha}\,\alpha}\,r^{\alpha} + \mathcal{O}(r^{2-\alpha})\,,}	&	\displaystyle{\mbox{for\, $r \to 0^+$}\,,} \vspace{0.1cm}\\
\displaystyle{e^{-\,\lambda\,r} \left( \sqrt{\pi \over 2}\;\lambda^{\alpha - 1/2}\, r^{-1/2} + \mathcal{O}\big(r^{-3/2}\big) \right),}	&	\displaystyle{\mbox{for\, $r \to +\infty$}\,.}
\end{array}\right. \label{asy0}
\end{gather}
The final outcome \cite[Cor. 2.6]{CoOd} is the very same family given in \eqref{eq: dom hamb alt}, with a slightly different domain representation\footnote{Note that in \cite[Eq. (2.22)]{CoOd} there is factor $ - 2 \pi  $ missing in the boundary condition.}:
\beq
	\label{eq: hamb action}
	\lf( H_{\alpha,0}^{(\beta)} + \la^2 \ri) \psi = \lf( H_{\alpha,0}^{(F)} + \la^2 \ri) \phi_{\lambda}\,,	
\eeq
\bml{
	\dom\big( H_{\alpha,0}^{(\beta)} \big) =  \lf\{ \psi \in L^2_{\mathrm{even}}(\R^2) \: \Big| \: \psi =  \phi_{\lambda} + q G_{\lambda}, \phi_{\lambda} \in \dom\big(H_{\alpha,0}^{(F)}\big)\,, \;  \ri. 	\\
	\lf.   q = \frac{2^{\alpha}\,\pi\, \Gamma(\alpha)}{\beta + c_{\alpha}\, \la^{2\alpha}}  \lim_{r \to 0^+} \frac{\alpha  \avg{\phi_{\lambda}}(r) + r\, \partial_r \avg{\phi_{\lambda}}(r)}{r^{\alpha}} \ri\},	 \label{eq: dom hamb}
}
Again, the limit case $\beta = + \infty$ is formally included understanding the boundary condition in \eqref{eq: dom hamb} as $q = 0$, which retrieves the Friedrichs realization $H_{\alpha,0}^{(F)}$\,.

The quadratic forms associated with the family of self-adjoint operators $ H_{\alpha,0}^{(\beta)} $ can be easily obtained by computing the expectation value of $ H_{\alpha,0}^{(\beta)} $ on functions in the relative domain $ \dom(H_{\alpha,0}^{(\beta)}) $. This yields the following expression
\beq
	\label{eq: form 0}
	Q_{\alpha,0}^{(\beta)}[\psi] = Q_{\alpha,0}^{(F)}[\phi_{\lambda}] - 2 \la^2 \Re \big[ q \braket{\phi_{\lambda}}{G_{\lambda}} \big] + \lf[ \beta + \lf( 1 - \alpha \ri) c_{\alpha} \la^{2\alpha} \ri] \lf| q \ri|^2,
\eeq
where $ Q_{\alpha,0}^{(\mathrm{F})} $ is the form associated with the Friedrichs extension $ H_{\alpha,0}^{(\mathrm{F})} $, {\it i.e.},
\beq
	Q_{\alpha,0}^{(F)} [\psi] = \!\int_{\mathbb{R}^2}\! d\mathbf{x}\; \lf| \lf(- i \nabla +  \mathbf{A}_{\alpha}  \ri) \psi \ri|^{2} \,,
\eeq
with domain given by the closure of $ C^{\infty}_{\mathrm{c}}(\R^2\setminus\{ \mathbf{0} \}) $ in the norm induced by the form itself. The quadratic forms $ Q_{\alpha,0}^{(\beta)} $ defined above are proven to be  \cite[Thm. 2.4]{CoOd} independent of $ \lambda $, closed and bounded from below on the domain
\beq
	\label{eq: dom form 0}
	\dom\big[Q_{\alpha,0}^{(\beta)}\big] = \lf\{ \psi \in L^2_{\mathrm{even}}(\R^2)  \: \Big| \: \psi =\phi_{\lambda} + q G_{\lambda}, \phi_{\lambda} \in \dom\big[Q_{\alpha,0}^{(F)}\big], \la > 0, q \in \mathbb{C} \ri\},
\eeq
which is notably independent of the parameter $ \beta \in \R $. Note also that in the form domain there is no boundary condition linking the ``charge'' $ q \in \mathbb{C} $ to the boundary value of the regular part $ \phi_{\lambda} $.

\subsection{Anyonic Schr\"{o}dinger operators for generic $ \mathbf{S} $, continuous at the origin}
\label{sec: regular}

We now deal with the modifications induced by the presence of $ \mathbf{S} $ to the family of quadratic forms \eqref{eq: form 0}. Since we aim at covering the most general magnetic potential, we do not make any assumption on the structure of $ \mathbf{S} $, but rather require some additional regularity close to the origin. More precisely, we assume in this section that, for certain $ r_0 > 0 $,
\beq
	\mathbf{S}\in
	\begin{cases}
		C^{0}(B_{r_{0}})\,, 	&	\mbox{if } \alpha \in (0,1/2)\,,		\\
		C^{0,\nu}(B_{r_0})\,, \mbox{ for some } \nu  > 2\alpha -1\,, & \mbox{if } \alpha \in [1/2,1)\,.
	\end{cases}
	\label{Shyp 2}
\eeq
Let us anticipate that the continuity hypothesis is necessary to ensure the well-posedness of the pointwise evaluation $\mathbf{S}(\mathbf{0})$, a key element in our arguments. As a matter of fact, this hypothesis is also sufficient for our constructions, when $0 < \alpha < 1/2$. On the contrary, we will need a stronger requirement (H\"{o}lder continuity) on the regularity of $ \mathbf{S}$ at the origin when $1/2 \leqslant \alpha < 1$. 

The model identified by the operator \eqref{HalS} and its self-adjoint realizations are naturally gauge invariant: given any regular function $ \varphi \in H^1(\R^2) $, the magnetic potentials $ \mathbf{S} $ and $ \mathbf{S} + \nabla \varphi $ generate the same physics, since the magnetic potential $ \mathbf{B} $, which is the only physically relevant quantity, is the same in both cases. In order to simplify the discussion, we make an explicit choice of the gauge and assume that $ \mathbf{S} $ is written in the Coulomb gauge, {\it i.e.}, 
\beq
	\label{eq: coulomb}
	\nabla \cdot \mathbf{S} = 0\,,
\eeq 
at least in distributional sense. Once the Coulomb gauge is chosen, the magnetic potential is still undefined by an additive constant and we choose such a constant so that
\begin{equation}\label{S0}
\mathbf{S}(\mathbf{0}) = \mathbf{0}\,,
\end{equation}
which can be obtained in any case via the global gauge transformation $\psi(\mathbf{x}) \to e^{-\, i \mathbf{S}(\mathbf{0})\,\cdot\, \mathbf{x}}\,\psi(\mathbf{x})$.
It is also useful to remark that, under these conditions, the operator \eqref{HalS} can be rewritten as
\begin{equation}\label{HalSC}
H_{\alpha,S} = \lf(- i \nabla + \mathbf{A}_{\alpha} \ri)^{2} + 2\,\mathbf{S} \cdot \lf( -i \nabla + \mathbf{A}_{\alpha} \ri) + \mathbf{S}^2\,,
\end{equation}
where the order in the cross term might be exchanged thanks to \eqref{eq: coulomb}.

We now focus on the other self-adjoint realizations of $ H_{\alpha,S} $ different from the Friedrichs one: the idea is to identify first the associated quadratic forms and, once such forms are proven to be closed and bounded from below, derive the explicit expression of the corresponding operators. Note indeed that the defect functions are not available for generic $ \mathbf{S} $ and therefore one can not perform the usual Von Neumann construction of the self-adjoint extensions nor apply the Kre\u{\i}n-Birman theory. In order to find the general expression of the quadratic forms, we thus start from an ansatz about their domain which recalls \eqref{eq: dom form 0}: we assume that each function $ \psi $ in the form domain (which as in \eqref{eq: dom form 0} is expected to be independent of the chosen extension) splits as a regular part $ \phi_{\lambda} \in \dom\big[Q_{\alpha,0}^{(F)}] $ plus a singular term proportional to a suitable defect function. Thanks to our assumption \eqref{Shyp 2}, we may decide that the behavior near the origin could still be given by $ G_{\lambda} $ introduced in \eqref{greenexp}. However, a possible divergence of $ \mathbf{S} $ at infinity would obviously call for a suitable modification of the defect function. Therefore, we introduce a smooth cut-off  $\chi : \mathbb{R}^2 \to [0,1] $, which is required to fulfill the following properties:
\beqn
	 &&\chi \in C^2_c(\R^2)\,; \nonumber \\
	 &&\chi(\xv) =  1\,,	\qquad	\mbox{for any } \xv \in B_{r_1}(\mathbf{0}), \mbox{for some } r_1 > 0\,. \label{eq: chi}
\eeqn 
Its explicit form may depend on the magnetic potential $ \mathbf{S} $ but we anticipate that, if $ \mathbf{S} \in L^{\infty}(\R^2) $, no cut-off is necessary, {\it i.e.}, one can take $ \chi \equiv 1 $. The new approximate defect function is then set equal to $ \chi G_{\lambda} $.

In order to derive the explicit expressions of the quadratic forms extending \eqref{QFr}, one can perform the following heuristic computation: by evaluating the expectation value of $ H_{\alpha,S}^{(F)} $ on a wave function of the form $ \psi = \phi_{\la} + q \chi G_{\la} $, with $ \phi_{\la} \in C^{\infty}_{\mathrm{c}}(\R^2\setminus\{\mathbf{0}\}) $, we formally get
\bdm
	\meanlrlr{\psi}{H_{\alpha,S}^{(F)}}{\psi} = \meanlrlr{\phi_{\la}}{H_{\alpha,S}^{(F)}}{\phi_{\la}} + 2\, \Re \lf[ q \meanlrlr{\phi_{\la}}{H_{\alpha,S}^{(F)}}{\chi G_{\la}} \ri] + \lf|q\ri|^2 \meanlrlr{ \chi G_{\la}}{H_{\alpha,S}^{(F)}}{\chi G_{\la}}.
\edm
Furthermore, since for $ \xv \neq \mathbf{0} $ (recall that the support of $ \phi_{\la} $ does not intersect the origin)
\bmln{
	H_{\alpha,S}^{(F)} \lf( \chi G_{\la} \ri)  = \chi H_{\alpha,S}^{(F)} G_{\la} + \lf( -\Delta \chi\ri) G_{\la} + 2\, (-i \nabla \chi) \cdot \lf(-i \nabla + \mathbf{A}_\alpha + \mathbf{S} \ri) G_\lambda
 \\
	=  \lf( \mathbf{S}^2 - \la^2 \ri) \chi G_{\la} + 2\, \chi  \mathbf{S} \cdot \lf( - i \nabla + \mathbf{A}_{\alpha} \ri) G_{\la} + \lf( -\Delta \chi\ri) G_{\la}+ 2 \,(-i \nabla \chi) \cdot \lf(-i \nabla + \mathbf{A}_\alpha + \mathbf{S} \ri) G_{\la}
}
two integrations by parts yield
\beq
 \label{QSheu}
\meanlrlr{\phi_{\la}}{H_{\alpha,S}^{(F)}}{\chi G_{\la}} = 2\, \braketr{ \lf(-i \nabla \!+\! \mathbf{A}_{\alpha}\ri)\,\phi_{\lambda}}{\lf(\mathbf{S}\,\chi - i\nabla \chi \ri)\,G_{\lambda}} + \braketr{ \phi_{\lambda}}{\lf( \mathbf{S}^2 \chi + \Delta \chi - \la^2 \chi \ri)\, G_{\lambda}} .
\eeq
Similarly, 
\bml{
	 \meanlrlr{ \chi G_{\la}}{H_{\alpha,S}^{(F)}}{\chi G_{\la}} =
\braketr{ \chi G_{\lambda}}{\lf( \mathbf{S}^2 \chi \!-\! \Delta \chi - \la^2 \chi \ri) G_{\lambda}}
+ 2\,  \braketr{\chi G_{\lambda}}{\lf(-i\nabla \chi \ri) \!\cdot\! \lf(-i\nabla \!+\! \mathbf{A}_{\alpha} \ri) G_{\lambda}}	\\
+ 2\, \braketr{\chi G_{\lambda}}{ \mathbf{S}\!\cdot\!\mathbf{A}_{\alpha}  \,\chi G_{\lambda}} \,.
}
Moreover, using the identity
		\bdm
			\lf\| \psi \ri\|^2_{2} = \lf\| \phi_{\la} \ri\|^2_{2} + 2\, \Re \lf[ q \braketr{\phi_{\la}}{G_{\la}} \ri] + \lf| q\ri|^2 \lf\| G_{\la} \ri\|^2_{2}\,,
		\edm
one can combine the terms proportional to $ \la^2 $, to recover the expression
\bml{
Q^{(\beta)}_{\alpha,S}[\psi] 
:= Q_{\alpha,S}^{(F)}[\phi_{\lambda}] -\,\lambda^2\,\|\psi\|_{2}^{2} + \lambda^2\, \|\phi_{\lambda}\|_{2}^{2} \\
 + 2 \,\Re \lf[q\,\lf(
2\,  \braketr{ \lf(-i \nabla \!+\! \mathbf{A}_{\alpha} \ri)\,\phi_{\lambda}}{\lf(\mathbf{S}\,\chi - i\nabla \chi\ri)\,G_{\lambda}}
+ \braketr{ \phi_{\lambda}}{\lf(\mathbf{S}^2 \chi + \Delta \chi \ri) G_{\lambda}} \ri) \ri]  \\
 + |q|^2\, \lf(\beta + c_{\alpha}\, \lambda^{2 \alpha} + \Xi_{\alpha,S}(\lambda) \ri)\;, \label{Qbeta}
}
where we have introduced the parameter $ \beta \in \R $ labelling the form and set
	\begin{equation}
		\Xi_{\alpha,S}(\lambda) := \braketr{ G_{\lambda}}{\big(\mathbf{S}^2\,\chi^2 -\chi\,\Delta \chi \big)\, G_{\lambda} }
		- \braketr{ G_{\lambda}}{ \nabla \chi^2 \cdot \nabla G_{\lambda} }
		+ 2\, \braketr{ G_{\lambda}}{\, \mathbf{S}\!\cdot\!\mathbf{A}_{\alpha}\,\chi^2\, G_{\lambda}}\,. \label{defXi}
	\end{equation}
In order to obtain such an expression, we got rid of the boundary terms coming from each integration by parts, which is going to be justified {\it a posteriori} (see the proof of \cref{cor:domHb1}). In particular, we have set  $\langle \chi G_{\lambda}\,|\, \mathbf{S} \!\cdot\! \nabla (\chi G_{\lambda})\rangle = 0$ and $\langle G_{\lambda}\,| \mathbf{A}_{\alpha} \!\cdot\! (\nabla \chi^2)\, G_{\lambda}\rangle = 0$, since formal integrations by parts yield (recall that $\chi$, $G_{\lambda}$ are real-valued, $G_{\lambda}$ is radial and, by \eqref{eq: coulomb}, $\nabla\!\cdot\!\mathbf{S} = 0$, besides $\nabla \cdot \mathbf{A}_{\alpha} = 0$)
\bdm	
	\braketr{\chi G_{\lambda}}{\mathbf{S} \cdot\!\nabla (\chi G_{\lambda})} = -\, \braketl{\mathbf{S} \cdot\!\nabla (\chi G_{\lambda})}{\chi G_{\lambda}} = -\, \braketr{\chi G_{\lambda}}{\mathbf{S} \cdot\!\nabla (\chi G_{\lambda})} \,,
\edm
	\bdm
		\braketr{ G_{\lambda}\,}{ \mathbf{A}_{\alpha} \cdot (\nabla \chi^2)\, G_{\lambda} }
		= -\,2\, \braketr{ G_{\lambda} \,}{\, \chi^2 \lf(\mathbf{A}_{\alpha} \cdot\! \nabla G_{\lambda}\ri) } = 0 \,.
	\edm

Once the form is obtained, it is then natural to identify its domain with the set of functions splitting in a regular part $ \phi_{\la} $ belonging to the domain of the quadratic form associated with the Friedrichs extension and a singular one proportional to the defect function, {\it i.e.}, 
\beq
 	\dom\big[Q^{(\beta)}_{\alpha,S}\big] := \lf\{\psi \!\in\! L_{\mathrm{even}}^2(\mathbb{R}^2)\,\big|\, \psi = \phi_{\lambda} + q\,\chi\,G_{\lambda} \,,\, \mbox{with } \phi_{\lambda} \!\in\! \dom\big[Q_{\alpha,S}^{(F)}\big],  \lambda \!>\! 0, q \!\in\! \mathbb{C} \ri\}\,.\label{Qbetadom}
\eeq

\begin{remark}[Well-posedness of $ Q^{(\beta)}_{\alpha,S} $]
	\label{rem:L2inner}
	\mbox{}	\\
	In Eq.\,\eqref{QSheu} and in the sequel the notation $\langle\,\cdot\,|\,\cdot\,\rangle$ indicates either the standard $L^2(\mathbb{R}^2)$ inner product or the duality pairing between suitable weighted spaces induced by it, which are in fact $ L^2 $ spaces over compact sets. With such a convention the expression of the quadratic form is well-defined (see \cref{sec: singular proof}). Let us give an example by looking at the first term on the second line of \eqref{Qbeta}.  We recall that $(-i \nabla \!+\! \mathbf{A}_{\alpha})\,\phi_{\lambda} \!\in\! L^2_{\mathrm{loc}}(\mathbb{R}^2)$ for all $\phi_{\lambda} \!\in\! \dom(Q_{\alpha,S}^{(F)})$, by \cref{rem:QF}, and that $G_{\lambda} \!\in\!L^2(\mathbb{R}^2)$ (see Eq.\,\eqref{G2norm}). On the other hand, since $ \mathbf{S} \!\in\!L^{\infty}_{\mathrm{loc}}(\mathbb{R}^2)$ and $\chi$ is twice differentiable with compact support, it appears that $\chi\, \mathbf{S} - i\nabla \chi$ is essentially bounded on $\mathbb{R}^2$ with support contained in $ \supp\,\chi$. Therefore, the expression $ \braketr{(-i \nabla + \mathbf{A}_{\alpha})\,\phi_{\lambda}}{ (\mathbf{S}\,\chi - i\nabla \chi)\,G_{\lambda}} $ should be meant as an inner product in $L^2(\supp\,\chi,d\mathbf{x})$ and, as such, it is well-posed. 
\end{remark}

We can now state our main result concerning the quadratic forms \eqref{Qbeta}.

	\begin{theorem}[Quadratic forms $ Q^{(\beta)}_{\alpha,S} $]
		\label{prop:Qbeta} 
		\mbox{}		\\
		Let $\alpha \in (0,1) $ and let $ \mathbf{S} $ satisfy \eqref{Shyp} and \eqref{Shyp 2}. Then, for any $\beta \in \mathbb{R}$, 
		 \ben[i)]
		 	\item the quadratic form $Q^{(\beta)}_{\alpha,S}$ defined in \eqref{Qbeta} is well-posed on the domain \eqref{Qbetadom} and independent of $ \lambda > 0 $ and the choice of $ \chi $;
		 	\item $Q^{(\beta)}_{\alpha,S}$ is also closed and bounded from below on the same domain;
		 	\item if $ \mathbf{S} \in L^{\infty}(\R^2) $ and $ \alpha \in (0,1/2) $, the following lower bound holds true
				\begin{equation}\label{lower1}
					{Q^{(\beta)}_{\alpha,S}[\psi] \over \|\psi\|_{2}^{2}} \geqslant - \left(\inf_{0 \,<\, \epsilon + 2 \eta \,<\, 1}\, \max\left\{ \lf\| \mathbf{S} \ri\|_{\infty}/\sqrt{\epsilon} \,,\,\lambda_{*}(\eta,\beta)\right\} \right)^{\!2}  \,,
				\end{equation}
				where $\lambda_{*}(\eta,\beta)  > 0$ is implicitly defined via
				\beq
					\label{lamst}
					\lambda_{*}^2 - \pi\,\alpha\,\tan(\pi\, \alpha) \,\| \mathbf{S} \|_{\infty}\,\lambda_{*} - {2\alpha\, \over \eta}\,\|S\|^2_{\infty} + {\beta \over c_{\alpha}}\, \lambda_{*}^{2\,(1-\alpha)} = 0\,.
				\eeq
		\een
	\end{theorem}
	
	\begin{remark}[Case $ \mathbf{S} \in L^{\infty}(\R^2) $]
		\label{rem: bounded}
		\mbox{}	\\
		The above result applies also to bounded magnetic potentials, but, whenever $ \mathbf{S} \in L^{\infty}(\R^2) $, it is possible to show that the expression of the quadratic form $Q^{(\beta)}_{\alpha,S}$ may be simplified a lot. Indeed, one can take $ \chi \equiv 1$, which leads to the form
		\bml{
			\widetilde{Q}^{(\beta)}_{\alpha,S}[\psi] := Q_{\alpha,S}^{(F)}[\phi_{\lambda}] 
- \lambda^2\,\|\psi\|_{2}^2 + \lambda^2\,\|\phi_{\lambda}\|_{2}^2
+ \lf\|\mathbf{S}\,\psi \ri\|_{2}^2 - \lf\| \mathbf{S} \,\phi_{\lambda} \ri\|_{2}^2 \\
+ 4\, \Re \big[ q \,\braketr{\lf(-i \nabla + \mathbf{A}_{\alpha} \ri)\phi_{\lambda}}{\mathbf{S}\, G_{\lambda}} \big]
+ |q|^2 \lf(\beta + c_{\alpha}\, \lambda^{2 \alpha} + \widetilde{\Xi}_{\alpha,S}(\lambda) \ri) ,\label{Qbetasmall}
		}
		\beq
			\widetilde{\Xi}_{\alpha,S}(\lambda) := 2\! \int_{\mathbb{R}^2}\!\! \diff\mathbf{x}\;  \mathbf{S} \!\cdot\! \mathbf{A}_{\alpha} \;G_{\lambda}^2\,, \label{defXismall}
		\end{equation}
		which can be proven to be an infinitesimally small perturbation of $ {Q}^{(\beta)}_{\alpha,S} $ and which is therefore closed and bounded from below on the same domain \eqref{Qbetadom}.
	\end{remark}
	
	\begin{remark}[Assumptions \eqref{Shyp} and \eqref{Shyp 2}]	
		\label{rem: regularity}
		\mbox{}	\\
		The regularity assumptions on the magnetic potential $ \mathbf{S} $ are in a sense optimal, namely it is hard to assume less regularity, without further hypothesis on the form of $ \mathbf{S} $, and still give a meaning to the expression \eqref{Qbeta} of the quadratic form. The key point is the well-posedness of the last term in \eqref{defXi}: the problem is that the function $ \chi^2\, \mathbf{S} \!\cdot\! \mathbf{A}_{\alpha} G^2_{\lambda} $ may fail to be integrable because of its local singularity at the origin. Indeed, in spite of the vanishing of $ \mathbf{S} $ at $ \mathbf{0} $, the function satisfies the following asymptotics
		\beq
			\label{eq: asympt 0}
			\lf( \mathbf{S} \cdot \mathbf{A}_{\alpha} G^2_{\la} \ri) (\xv) \underset{\xv \to \mathbf{0}}{\sim} \frac{\lf| \mathbf{S}(\xv) \ri|}{\lf| \xv \ri|^{1+2\alpha}},
		\eeq
		so that, if $ \alpha \geq 1/2 $, unless $ \mathbf{S} $ goes to zero fast enough, that term in the form blows up. On the other hand, H\"{o}lder continuity assumed in \eqref{Shyp 2} implies that $ |\mathbf{S}(\mathbf{x})| \leqslant C\,|\mathbf{x}|^{\nu} $, for $ \nu > 2 \alpha - 1 $,  which is almost\footnote{One might in fact relax a bit the assumption for $ \alpha \in [1/2,1) $ and require that $ \mathbf{S} $ is such that the r.h.s. of \eqref{eq: asympt 0} is integrable. This however would complicate certain estimates in the proofs and therefore we stick to the assumption \eqref{Shyp 2}.} the minimal request on the rate of convergence to ensure local integrability of the function above. Note indeed that a quick inspection of the form reveals that the only divergence, which may occur under the assumption \eqref{Shyp}, is the one we are discussing, and therefore dropping \eqref{Shyp 2} would result in an obstruction to define the extended quadratic forms as in \eqref{Qbeta}. A different scheme for a particular singularity of the type just mentioned is described in the subsequent \cref{sec: discontinuous}.
	\end{remark}
	
	\begin{remark}[Lower bound \eqref{lower1}]
		\label{rem: lower bound}
		\mbox{}	\\
		The lower bound provided by  \eqref{lower1} is rather implicit but it is possible to show that
		\beqn
			\lambda_{*}  \leqslant &{\pi \over 2}\,\alpha \tan(\pi\,\alpha) \lf\| \mathbf{S} \ri\|_{\infty} \left(1 + \sqrt{1 \!+\! {8 \over \pi^2\eta\,\alpha \tan(\pi\,\alpha)}}\right),  &		\mbox{if } \beta \!\geqslant\! 0\,, \nonumber	\\
			\lambda_{*} >  & {\pi \over 2}\,\alpha\,\tan(\pi\,\alpha)\,\lf\| \mathbf{S} \ri\|_{\infty} \left(1 + \sqrt{1 + {8 \over \pi^2\eta\,\alpha\,\tan(\pi\,\alpha)}}\right),  &	\mbox{if }  \beta < 0\,, \nonumber
		\eeqn
		which allows to extract a more explicit bound in \eqref{lower1}. Note that the bound from below is always negative, although one can find certain conditions on $ \beta $ and $ \mathbf{S} $ to get a positive quadratic form ({\it e.g.}, $ \beta \geq 0 $ and $ \mathbf{S} \cdot \mathbf{A}_{\alpha} \geq 0 $ pointwise). In general, however, even in the simple case described in \cref{rem: bounded} and for $ \beta \geq 0 $, the term $ \widetilde{\Xi}_{\alpha,S}(\lambda) $ in \eqref{Qbetasmall} (and the corresponding one in \eqref{Qbeta}) has no given sign and scales as $ \lambda^{2\alpha - 2} $ when $ \lambda \to 0^+ $, so that it can not be controlled by the positive term $ c_{\alpha} \lambda^{2\alpha} $. Finally, we point out that a bound similar to \eqref{lower1} can in principle be proven also without the boundedness assumption on $ \mathbf{S} $, but in that case it is even more implicit and thus we omit it for the sake of brevity.
	\end{remark}

	\begin{remark}[Electric potentials]
		\label{rem: electric}
		\mbox{}		\\
		The above construction may be extended also in presence of electric potentials ({\it e.g.}, trapping or interaction potentials), which are regular enough close to the origin, in the same spirit of \cite[\textsection 2.2]{CoOd}. We skip the discussion for the sake of brevity.
	\end{remark}
	
The operator family associated with the quadratic forms $ Q_{\alpha,S}^{(\beta)} $ is described in details in next

	\begin{corollary}[Self-adjoint extensions $ H^{(\beta)}_{\alpha,S} $]
		\label{cor:domHb1}
		\mbox{}	\\
		Under the same assumptions of \cref{prop:Qbeta}, the operators $ H^{(\beta)}_{\alpha,S} $, $ \beta \in \R$, associated with the quadratic forms $Q^{(\beta)}_{\alpha,S}$, which are given by
		\bml{
			\label{domHbe1}
			\dom \big(H^{(\beta)}_{\alpha,S}\big) 
			= \lf\{ \psi = \phi_{\lambda} \!+\! q\,\chi\,G_{\lambda} \!\in\! 								 			\dom \big[Q^{(\beta)}_{\alpha,S}\big] \,\Big|\,
H_{\alpha,S}\, \phi_{\lambda} + 2\,q\, \chi\, \mathbf{S} \!\cdot\!(-i\nabla\!+\!\mathbf{A}_{\alpha}) G_{\lambda} \in L^2(\mathbb{R}^2)\,, \ri.	\\
			\lf. q = {2^{\alpha}\, \pi\,\Gamma(\alpha) \over \beta + c_{\alpha}\, 					\lambda^{2 \alpha}}\; \lim_{r \to 0^+}\!  \frac{\alpha\, \avg{ \phi_{\lambda}}(r) + r\,\partial_r \avg{\phi_{\lambda}}(r)}{r^{\alpha}} \ri\}\,,
			}
			\bml{
				\label{Hbe1}
				H^{(\beta)}_{\alpha,S}\, \psi  = H_{\alpha,S}\,\phi_{\lambda}
				+ q \lf[ \lf( \mathbf{S}^2\! - \lambda^2 \ri)\,\chi\,G_{\lambda}\!
				+ 2\,\chi\,\mathbf{S} \!\cdot\! \lf(-i\nabla \!+\! \mathbf{A}_{\alpha} \ri) G_{\lambda}\! \ri. \\
				\lf. + 2\,(-i \nabla \chi)\!\cdot\! \lf(-i \nabla\!+\!\mathbf{A}_{\alpha}\!+\!\mathbf{S} \ri) G_{\lambda}\!
				- (\Delta \chi) G_{\lambda} \ri]\,,
			}
			identify a one-parameter family of self-adjoint extensions of $ \dot{H}_{\alpha,S} $ in $ L^2(\mathbb{R}^2) $. In particular, if $ \mathbf{S} \in L^{\infty}(\R^2) $, this family parametrizes all anyonic self-adjoint realizations of $ H_{\alpha,S} $ in $ L_{\mathrm{even}}^2(\R^2) $.
	\end{corollary}

\begin{remark}[Operator domain (I)]
	\label{rem: domain I}
	\mbox{}	\\
	The representation \eqref{domHbe1} of the operator domain provides a simple characterization of the charge $q\!\in\!\mathbb{C}$ in terms of the asymptotics at the origin of the regular part $\phi_{\lambda}$ of the wave function in the form domain: as it is typical for singular perturbations, in the operator domain the coefficient of the defect function is determined via a boundary condition, which is absent in the form domain. However, \eqref{domHbe1} also contains the somehow unusual requirement $H_{\alpha,S} \phi_{\lambda} + 2\,q\, \chi\, \mathbf{S} \cdot (-i\nabla\!+\!\mathbf{A}_{\alpha}) G_{\lambda} \!\in\! L^2(\mathbb{R}^2)$, which has unexpected implications, since $ \phi_{\la} $ {\it does not} belong to the domain of the Friedrichs extension $ H_{\alpha,S}^{(F)} $ (though it is in the Friedrichs {\it form} domain). In particular, let us stress that $H_{\alpha,S}\,\phi_{\lambda}$ and $ \mathbf{S} \cdot (-i\nabla + \mathbf{A}_{\alpha})G_{\lambda}$ are not in general  separately square-integrable, due to their singular behaviour at the origin. This inconvenience is due to the use of $G_{\lambda}$, namely the defect function for ${H}_{\alpha,0}$, rather than the true defect function for ${H}_{\alpha,S}$. 
\end{remark}

	\begin{remark}[Operator domain (II)]
		\label{rem: domain II}
		\mbox{}	\\
		The operator domain representation may be simplified, if the magnetic potential is more regular than what we have assumed. For instance, if we require that $ \mathbf{S} $ is Lipschitz in a neighborhood of the origin, {\it i.e.}, $ \mathbf{S} \in C^{0,\nu}(B_{r_0})$, for some $r_{0} \!>\! 0$ and $ \nu > \alpha $, it can be easily checked that the weird behavior described in the previous \cref{rem: domain I} can not occur and $\mathbf{S}  \cdot (-i\nabla + \mathbf{A}_{\alpha}) G_{\lambda} \!\in\! L^2(\mathbb{R}^2)$. In this case, the operator domain can be rewritten as
		\bml{
			\label{domHbe2}
			\dom \big(H^{(\beta)}_{\alpha,S}\big) 
			= \lf\{ \psi = \phi_{\lambda} \!+\! q\,\chi\,G_{\lambda} \!\in\! 								 			\dom \big[Q^{(\beta)}_{\alpha,S}\big] \,\Big|\,
\phi_{\la} \in \dom \big( H_{\alpha,S}^{(F)} \big)\,, \ri.	\\
			\lf. q = {2^{\alpha}\, \pi\,\Gamma(\alpha) \over \beta + c_{\alpha}\, 					\lambda^{2 \alpha}}\; \lim_{r \to 0^+}\!  \frac{\alpha\, \avg{ \phi_{\lambda}}(r) \!+\! r\,\partial_r \avg{\phi_{\lambda}}(r)}{r^{\alpha}} \ri\}\,.
			}
			Let us also mention that, at least in the anyonic context, again under the assumption $ \mathbf{S} \in C^{0,\nu}(B_{r_0})$, with $ \nu > \alpha $, it is possible to characterize the operator domain in a way analogous to \eqref{eq: dom hamb alt}, namely,
	\bml{
		\dom\big( H_{\alpha,S}^{(\beta)} \big) =  \lf\{ \psi \in L^2_{\mathrm{even}}(\R^2) \: \Big| \: H_{\alpha,S}\, \psi \in L^2_{\mathrm{even}}(\R^2)\,, \;  \ri. 	\\
		\lf. \avg{\psi}(r) \underset{r \to 0^{+}}{\sim} C \left(r^{\alpha} + {2^{2\alpha}\, \pi\, \alpha\,\Gamma^2(\alpha) \over \beta}\,r^{-\alpha}\right) + o(r)  \;\mbox{for some } C \in \R \ri\}.
		\label{domHbe2 alt}
	}

\end{remark}

\begin{remark}[Other self-adjoint extensions]
	\label{rem: sa extensions}
	\mbox{}	\\
	We do not know whether the family of operators $ \big\{ H_{\alpha,S}^{(\beta)} \big\}_{ \beta \in \mathbb{R} } $ exhausts all possible self-adjoint extensions of the symmetric operator $ \dot{H}_{\alpha,S} $ in the full generality of hypothesis \eqref{Shyp}. This is highly expected in the case of anyons. Indeed, Corollary \ref{cor:domHb1} provides this kind of result assuming in addition that $ \mathbf{S} $ is bounded also at infinity. We believe that the latter assumption is actually unnecessary, since the deficiency indices of $ \dot{H}_{\alpha,S} $ should reasonably depend just on the local behaviour of $ \mathbf{S} $ close to the Aharonov-Bohm singularity at $ \mathbf{x} = \mathbf{0} $.
	On the other hand, note that more self-adjoint extensions would certainly appear, if one removed the symmetry restriction to even functions (see again \cref{prop: special}), {\it e.g.}, for a model describing a particle moving in a Aharonov-Bohm magnetic field. In this case one should expect to observe the emergence of another defect function living in the subspace with angular momentum $ -1 $. Such extensions might be found by constructing the corresponding quadratic forms, as we did for the rotationally symmetric ones, but their implementation would require more assumptions on the magnetic potential $ \mathbf{S} $ and make the analysis much more involved. We thus skip this discussion.
\end{remark}

\subsection{Anyonic Schr\"{o}dinger operators for generic $ \mathbf{S} $, with a specific type of discontinuity}
\label{sec: discontinuous}

By a variation of the arguments described in the previous section it is possible to characterize a family of self-adjoint realizations of $ H_{\alpha,S} $ also for magnetic perturbations $ \mathbf{S} $ which are singular at the origin, where the Aharonov-Bohm flux is located. More precisely, in addition to the basic hypothesis \eqref{Shyp}, making reference to the decomposition 
\beq
	\mathbf{S}(\mathbf{x}) = S_{\perp}(\mathbf{x})\,\hat{\mathbf{x}}^{\perp} + S_{\parallel}(\mathbf{x})\,\hat{\mathbf{x}}\,, \qquad S_{\perp}(\mathbf{x}) := \mathbf{S}(\mathbf{x}) \cdot \hat{\mathbf{x}}^{\perp}\,, \quad S_{\parallel}(\mathbf{x}) := \mathbf{S}(\mathbf{x}) \cdot \hat{\mathbf{x}}\,,
\eeq
we assume\footnote{In fact, the condition on $ S_{\parallel} $ may be relaxed a bit  in the same spirit of the previous section and one could simply require $ \nu > 2 \alpha -1 $ (recall that $ 2 \alpha  - 1 < \alpha $, since $ \alpha < 1 $), since the well-posedness of the quadratic form requires such an assumption. However, this would result in a more complicate expression of the operator domain and action (compare with \cref{rem: domain I}). Therefore, we stick to the condition $ \nu > \alpha $ for the sake of simplicity.} in this section that, for certain $r_0 > 0$,
\beq
	\label{Shyp 2a}
	S_{\perp}, S_{\parallel} \in
	\begin{cases}
		C^{0}(B_{r_{0}})\,, 	&	\mbox{if } \alpha \in (0,1/2)\,,		\\
		C^{0,\nu}(B_{r_0})\,, \mbox{ for some } \nu > \alpha \,, & \mbox{if } \alpha \in [1/2,1)\,
	\end{cases}
\eeq
and
\beq
	\lim_{\xv \to 0} S_{\parallel}(\xv) = 0.
\eeq

It is important to remark that the second one of the assumptions \eqref{Shyp 2a}, {\it i.e.}, the vanishing of the parallel component of $ \mathbf{S} $ at the origin, is actually needed in order to make the choice of the gauge, which is assumed to be Coulomb's one in this section too, {\it i.e.}, in distributional sense,
\bdm
	\nabla \cdot \mathbf{S} = 0.
\edm
Indeed, once the above condition is written in terms of the components of $ \mathbf{S} $, one gets
\bdm
	\nabla \cdot \mathbf{S} = \nabla S_{\parallel} \cdot \hat{\xv} + \nabla S_{\perp} \cdot \hat{\xv}^{\perp} + \frac{S_{\parallel}}{|\xv|},
\edm
and, even for smooth components, the last term is the only one diverging at the origin, unless $ S_{\parallel} $ vanishes there. Hence, in order for the Coulomb gauge condition to be satisfied at least in distributional sense, we have to assume that $ S_{\parallel}(\xv) \to 0  $, as $ \xv \to 0 $. In fact, if $ S_{\parallel} $ was regular enough, one could set $ S_{\parallel}(0) = 0 $ via a direct gauge transformation. 

In this section only we restrict the attention to the case where
\beq
	S_{\perp}(0) \neq 0\,,
	\label{Shyp 2b}
\eeq
since otherwise $ \mathbf{S} \in C^{0}(B_{r_0}) $ and we could refer to the analysis of the previous section. Indeed, \eqref{Shyp 2a} and \eqref{Shyp 2b} imply that $\mathbf{S}(\mathbf{x})$ admits no limit for $ \mathbf{x} \to \mathbf{0} $, which makes $\mathbf{S}$ discontinuous at the origin. As a consequence, we can not implement the condition \eqref{S0} via a gauge transformation as we did in the preceding section. It should also be noted that \eqref{Shyp 2a} yields
\beq \label{SAlim}
	\mathbf{S} \cdot \mathbf{A}_{\alpha}  
\sim  {\alpha\, S_{\perp}(\mathbf{0}) \over |\mathbf{x}|}, \qquad \mbox{as } |\mathbf{x}| \to 0^+ \,,
\eeq
which, in view of the expansion \eqref{asy0} for $G_{\lambda}$, makes apparent that the term $ \braketr{ G_{\lambda}}{\, \mathbf{S}\!\cdot\!\mathbf{A}_{\alpha}\,\chi^2\, G_{\lambda}} $ in \eqref{defXi} is actually divergent for $\alpha \in [1/2,1)$. So, the very expression \eqref{Qbeta} for the perturbed quadratic form $ Q^{(\beta)}_{\alpha,S} $ discussed previously in \cref{sec: regular} is ill-defined in the present context.

Taking into account the above considerations, in the sequel we develop an alternative approach to construct self-adjoint realizations of $ H_{\alpha,S} $ different from the Friedrichs one. This approach relies on a modification of the defect function $ G_{\la} $, changing the next-to-leading order term in the asymptotics as $ |\xv| \to 0^+ $. To this avail, we introduce the radial deformation function
\begin{equation}\label{eq:defzetan}
\zeta_{\alpha}(r) := 
	\begin{cases}
		1\,, &	\mbox{for }  0 < \alpha < 1/2,  \\
		\displaystyle{1 + S_{\perp}(\mathbf{0})\,r\,\log r\,,} &	 \mbox{for }  \alpha = 1/2, \\
		\displaystyle{1 - {\alpha\, S_{\perp}(\mathbf{0}) \over \alpha -1/2}\,r\,,}	&									 \mbox{for } 1/2 < \alpha < 1.
	\end{cases} 
\end{equation}
We anticipate that the singular behaviour of the above expressions close to the origin is key to compensate the previously mentioned divergences related to the mixed product $\mathbf{S}\!\cdot\!\mathbf{A}_{\alpha}$. On the other hand, to deal with possible divergences of $ \mathbf{S} $ at infinity we introduce again a smooth cut-off $\chi : \mathbb{R}^2 \to [0,1] $ as in \eqref{eq: chi}. 

Taking $  \chi\,\zeta_{\alpha} G_{\la}  $ as a defect function and decomposing the wave functions in the form domain as $ \psi = \phi_{\la} + q\, \chi\,\zeta_{\alpha} G_{\la} $, with $ \phi_{\la} \in C^{\infty}_{\mathrm{c}}(\R^2\setminus\{\mathbf{0}\}) $ and $q \in \mathbb{C}$, by heuristic computations similar to those described in \cref{sec: regular}, we are led to consider the family of quadratic forms
\bml{
Q^{(\beta)}_{\alpha,S}[\psi] 
	:= Q_{\alpha,S}^{(F)}[\phi_{\lambda}] -\,\lambda^2\,\|\psi\|_{2}^{2} + \lambda^2\, \|\phi_{\lambda}\|_{2}^{2}  \\
	+ 2 \,\Re \lf[ q \Big(
		2 \braketr{ \lf(-i \nabla \!+\! \mathbf{A}_{\alpha}\ri)\phi_{\lambda}}{\big(\mathbf{S}\,\chi \zeta_{\alpha}\! + \zeta_{\alpha} (- i\nabla \chi) + \chi (- i\nabla \zeta_{\alpha}) \big)\, G_{\lambda}} \ri. \\
		\lf. + \braketr{ \phi_{\lambda}}{\big( \mathbf{S}^2\, \chi \zeta_{\alpha}\! + \zeta_{\alpha}\,\Delta \chi + 2\,\nabla \chi \!\cdot\! \nabla\zeta_{\alpha} \big)\, G_{\lambda}}
		+ \braketr{ \phi_{\lambda}}{\,\chi\,\Delta \zeta_{\alpha}\, G_{\lambda}} 
		\Big) \ri] \\
+ |q|^2 \lf(\beta + c_{\alpha}\, \lambda^{2 \alpha} + \Xi_{\alpha}(\lambda) \ri), \label{Qbeta2}
}
where $\beta \in \mathbb{R}$ is the extension parameter labelling the form and we have set
\bml{\label{defXi2}
\Xi_{\alpha}(\lambda) 
	:= \braketr{G_{\lambda}}{\,\mathbf{S}^2\, \chi^2\, \zeta_{\alpha}^2\, G_{\lambda}}
		- \braketr{G_{\lambda}}{\, \zeta_{\alpha}^2\, \nabla \chi^2  \cdot \nabla G_{\lambda} } 
		- \braketr{G_{\lambda}}{\, \zeta_{\alpha}^2\, \chi\,\Delta \chi \, G_{\lambda}} 
		- \braketr{G_{\lambda}}{\, \zeta_{\alpha}\, \big(\nabla \chi^2 \!\cdot \!\nabla \zeta_{\alpha}\big)\, G_{\lambda}}\\
		+ \braketr{G_{\lambda}}{\,\chi^2 \zeta_{\alpha} \lf(2\,\mathbf{S}\!\cdot\!\mathbf{A}_{\alpha}\, \zeta_{\alpha}\, G_{\lambda} - \Delta \zeta_{\alpha}\,  G_{\lambda} - 2\,\nabla \zeta_{\alpha} \cdot\! \nabla G_{\lambda} \ri)}\,.
}
The natural domain of definition of the form \eqref{Qbeta2} is
\beq
 	\dom\big[Q^{(\beta)}_{\alpha,S}\big] := \lf\{\psi \!\in\! L_{\mathrm{even}}^2(\mathbb{R}^2)\,\big|\, \psi = \phi_{\lambda} + q\,\chi\,\zeta_{\alpha} G_{\lambda} \,,\, \mbox{with } \phi_{\lambda} \!\in\! \dom\big[Q_{\alpha,S}^{(F)}\big],  \lambda \!>\! 0, q \!\in\! \mathbb{C} \ri\}\,.\label{Qbetadom2}
\eeq

\begin{remark}[Well-posedness of $ Q^{(\beta)}_{\alpha,S} $]
	\label{rem:L2inner2}
	\mbox{}	\\
	Regarding the meaning of the notation $\langle\,\cdot\,|\,\cdot\,\rangle$ in \eqref{Qbeta2} and \eqref{defXi2}, one can make considerations analogous to those reported in \cref{rem:L2inner}, {\it i.e.}, it may stand either for the conventional inner product in $ L^2(\R^2) $ or, more in general, for a pairing between weighted spaces. With such a convention the expression \eqref{Qbeta2} is well-posed.
\end{remark}

The analogues of \cref{prop:Qbeta} and \cref{cor:domHb1} read as follows.

\begin{theorem}[Quadratic forms $ Q^{(\beta)}_{\alpha,S} $]
		\label{prop:Qbeta2} 
		\mbox{}		\\
		Let $\alpha \in (0,1) $ and let $ \mathbf{S} $ satisfy \eqref{Shyp} and \eqref{Shyp 2a}. Then, for any $\beta \in \mathbb{R}$, 
		 \ben[i)]
		 	\item the quadratic form $Q^{(\beta)}_{\alpha,S}$ defined in \eqref{Qbeta2} is well-posed on the domain \eqref{Qbetadom2} and independent of $ \lambda > 0 $ and the choice of $ \chi $;
		 	\item $Q^{(\beta)}_{\alpha,S}$ is also closed and bounded from below on the same domain.
		\een		
	\end{theorem}
	
	\begin{remark}[Lower bound]
		\mbox{}	\\
		As in \cref{prop:Qbeta}, it is not difficult to exploit the argument to prove the boundedness from below of the quadratic form to derive an explicit lower bound on the form. However, because of the arbitrarity of the cut-off function \eqref{eq:defzetan} (any other function with the same asymptotics at the origin would indeed work), such a bound is not so meaningful and thus we omit it.
	\end{remark}
	
	We are now in position to state the result about the self-adjoint extensions of $ \dot{H}_{\alpha,S} $.

\begin{corollary}[Self-adjoint extensions $ H^{(\beta)}_{\alpha,S} $]
		\label{cor:domHb12}
		\mbox{}	\\
		Under the same assumptions of \cref{prop:Qbeta2}, the operators $ H^{(\beta)}_{\alpha,S} $, $ \beta \in \R$, associated with the quadratic forms $Q^{(\beta)}_{\alpha,S}$, which are given by
		\bml{
			\label{domHbe12}
			\dom \big(H^{(\beta)}_{\alpha,S}\big) 
			= \lf\{ \psi = \phi_{\lambda} \!+\! q\,\chi\,G_{\lambda} \!\in\! \dom \big[Q^{(\beta)}_{\alpha,S}\big] \,\Big|\,\ri. \phi_{\lambda} \in \dom \big(H^{(F)}_{\alpha,S}\big),  \\
			 \lf. q = { 2^{\alpha}\, \pi\,\Gamma(\alpha)\over \beta + c_{\alpha}\, \lambda^{2 \alpha}}\; \lim_{r \to 0^+}  \frac{\alpha \avg{ \phi_{\lambda}}\!(r) + r\,\partial_r \!\avg{\phi_{\lambda}}\!(r)}{r^{\alpha}} \ri\}\,,
			}
			\bml{
				\label{Hbe12}
				H^{(\beta)}_{\alpha,S}\, \psi  = H_{\alpha,S}\,\phi_{\lambda}
				+ q \lf[ \lf( \mathbf{S}^2\! - \lambda^2 \ri)\,\chi\,\zeta_{\alpha}\,G_{\lambda}\!
				+ \,2\,\chi\,\zeta_{\alpha}\,\mathbf{S} \!\cdot\! \lf(-i\nabla \!+\! \mathbf{A}_{\alpha} \ri) G_{\lambda}\! \ri. \\
				\lf. + \,2\,\zeta_{\alpha} (-i \nabla \chi) \cdot (-i \nabla\!+\!\mathbf{A}_{\alpha}\!+\!\mathbf{S}) G_{\lambda} + 2 \,\chi (-i \nabla \zeta_{\alpha}) \cdot (-i \nabla\!+\!\mathbf{S})\, G_{\lambda} \ri. \\
				\lf. - \,(\zeta_{\alpha}\,\Delta \chi + 2\, \nabla \chi \cdot \nabla \zeta_{\alpha} + \chi\, \Delta\zeta_{\alpha})\, G_{\lambda} \ri]\,,
			}
			identify a one-parameter family of self-adjoint extensions of $ \dot{H}_{\alpha,S} $ in $ L^2(\mathbb{R}^2) $. In particular, if $ \mathbf{S} \in L^{\infty}(\R^2) $ this family parametrizes all anyonic self-adjoint realizations of $ H_{\alpha,S} $ in $ L_{\mathrm{even}}^2(\R^2) $.
	\end{corollary}

\begin{remark}[Operator domain]
	\mbox{}	\\
	Under the assumptions \eqref{Shyp 2a} on the magnetic potential $ \mathbf{S} $, the r.h.s. of \eqref{Hbe12} is automatically in $ L^2(\R^2) $, because, thanks to the choice of $ \zeta_{\alpha} $ as well as the regularity of $ S_{\perp}$  and $ S_{\parallel} $ close to the origin, all the leading order divergences there cancel out. 
Let us further remark that, similarly to what we pointed out in \cref{rem: domain II} (see especially \eqref{domHbe2 alt} and the related comments), it is possible to obtain a more explicit representation of the operator domain, characterizing the asymptotic behaviour of its elements close to the origin in the spirit of \cite{BDG,DR,DR18,DR20} (see also next \cref{sec: special}). More precisely, in the anyonic setting the angular average of any $\psi \in \dom\big( H_{\alpha,S}^{(\beta)} \big)$ behaves\footnote{To make a comparison with \cite{DR20}, let us highlight that the asymptotic expansions in \eqref{asy Sperp} lead to the following indentifications of the parameters $m \equiv m_{\mbox{{\tiny \cite{DR20}}}}, \beta \equiv \beta_{\mbox{{\tiny \cite{DR20}}}}, \kappa \equiv \kappa_{\mbox{{\tiny \cite{DR20}}}}, \nu \equiv \nu_{\mbox{{\tiny \cite{DR20}}}}$ of the cited reference (see, especially, \S 2.4 therein): $
m_{\mbox{{\tiny \cite{DR20}}}} = \alpha $, $ \beta_{\mbox{{\tiny \cite{DR20}}}}= -\, 2 \alpha\,S_\perp(\mathbf{0}) $, $ \kappa_{\mbox{{\tiny \cite{DR20}}}} = {2^{2\alpha}\, \pi\, \alpha\,\Gamma^2(\alpha) \over \beta} $ and $ \nu_{\mbox{{\tiny \cite{DR20}}}} = {\beta \over \pi^2} $.
} as follows for a suitable $C \in \R$:
\begin{equation}
\avg{\psi}(r) \underset{r \to 0^+}{\sim}  \left\{\!\begin{array}{ll}
\displaystyle{C \left( r^{\alpha} + {2^{2\alpha}\, \pi\, \alpha\,\Gamma^2(\alpha) \over \beta}\,r^{-\alpha}\right) + o(r^{1-\alpha})\,,} & \displaystyle{\mbox{for $0 < \alpha < 1/2$\,,}} \vspace{0.1cm}\\
\displaystyle{C \left( {1 \over \sqrt{r}}\, \big(1 + S_{\perp}(\mathbf{0})\, r \log r)\big) + {\beta \over \pi^2}\,r \right) + o(r)\,,} & \displaystyle{\mbox{for $\alpha = 1/2$\,,}} \vspace{0.1cm} \\
\displaystyle{C \left( r^{\alpha} + {2^{2\alpha}\, \pi\, \alpha\,\Gamma^2(\alpha) \over \beta}\,r^{-\alpha}\,\Big(1 - {\alpha\, S_{\perp}(\mathbf{0}) \over \alpha - 1/2}\,r\Big) \right) + o(r)\,,} & \displaystyle{\mbox{for $1/2 < \alpha < 1$\,.}}
\end{array}\right.
\label{asy Sperp}
\end{equation}

\end{remark}

\subsection{Aharonov-Bohm Schr\"{o}dinger operators for special $ \mathbf{S} $}
\label{sec: special}

In this section we take a different view point and, instead of considering the most general magnetic potential $ \mathbf{S} $, we focus on potentials of a specific form, although still satisfying \eqref{Shyp}. We thus assume that, for certain $ r_0 > 0 $,
\beq
	\label{Shyp 3}
	\mathbf{S}(\mathbf{x}) = S\lf(|\mathbf{x}|\ri)\, \hat{\mathbf{x}}^{\perp},  \qquad \mbox{for some } S \in C^{0,1}\big([0,r_0)\big) \cap L^{\infty}(\mathbb{R}_{+}). 
\end{equation}
Note that such a potential can not be continuous at the origin (and then satisfy assumption \eqref{Shyp 2}), unless $ S(0) = 0 $. Actually, if this is the case, one can easily realize that $ \mathbf{S} \in C^{0}(B_{r_0}) $. Therefore, the most interesting case $ S(0) \neq 0 $, which is not included, is fact covered by the discussion in \cref{sec: discontinuous}. Note also that, for a generic vector potential $ \mathbf{S}(\mathbf{x})$, the decomposition $ \mathbf{S}(\mathbf{x}) = S_{\parallel}(\mathbf{x})\, \hat{\mathbf{x}} + S_{\perp}(\mathbf{x})\,\hat{\mathbf{x}}^{\perp}$ holds for any $\mathbf{x} \in \mathbb{R}^2 \setminus \{\mathbf{0}\}$\,. Moreover, if $ S_{\parallel} $ is {\it radial} and regular enough, one can get rid of the radial part of $ \mathbf{S} $, via the gauge transformation  $\psi(\mathbf{x}) \to e^{-\, i \varphi(|\xv|)}\,\psi(\mathbf{x})$, with, {\it e.g.},
\bdm
	\varphi(|\xv|) : = \int_0^{|\xv|} \diff r \: S_{\parallel}(r),
\edm
since $ \nabla \varphi = S_{\parallel}(\mathbf{x})\, \hat{\mathbf{x}} $. Notice also that the Coulomb gauge imposes no restriction on $ S $, since the condition $\nabla \cdot \mathbf{S} = 0$ is identically fulfilled in the sense of distributions.

Before stating the main result of the section, we present a special case, which can be solved explicitly: assume that $ S(|\xv|) \equiv S(0) \neq 0 $ constant, then applying the cylindrical harmonics decomposition
\begin{equation}\label{RepSph}
L^2(\mathbb{R}^2) = \bigoplus_{k \in \mathbb{Z}}\, L^2(\mathbb{R}_{+},r\, dr) \otimes \mbox{span}\left({e^{i k \theta} \over \sqrt{2\pi}}\right) , \qquad
\psi(\mathbf{x}) = \sum_{k \in \mathbb{Z}} \psi_{k}(r)\,{e^{i k \theta} \over \sqrt{2\pi}}\;,
\eeq
one gets
\begin{equation}
	H_{\alpha,S} -\mathbf{S}^2(0) = \sum_{k \in \mathbb{Z}} V^{-1}\, L_{\alpha,S}^{(k)}  \, V \otimes \mathbf{1}\;, \label{HspL}
\end{equation}
where we introduced the unitary transformation $V : L^2(\mathbb{R}_{+},r\, dr) \to L^2(\mathbb{R}_{+}, dr)$, $(V\, \psi_k)(r) := \sqrt{r}\,\psi_k(r)$ and the family of radial differential operators
\beq
	\label{eq: Lalpha}
L_{\alpha,S}^{(k)} := -\,{d^2\, \over dr^2} + {(k + \alpha)^2 - 1/4 \over r^2} + {2(k + \alpha)S(0) \over r}\,, 
\eeq
a.k.a. {\it Whittaker operators}. A comprehensive analysis of all self-adjoint realizations on $L^2(\mathbb{R}_{+}, dr)$ of the differential operator $L_{\alpha,\xi}^{(k)}$, for any $k \!\in\! \mathbb{Z}$, is performed\footnote{In order to simplify the comparison, we point out that the parameters $\alpha \equiv m^2$ and $\beta$ used in \cite{DR18} and \cite{DR20} (hereafter denoted with $\alpha_{\mbox{{\tiny\cite{DR20}}}} \equiv m_{\mbox{{\tiny\cite{DR20}}}}^2$ and $\beta_{\mbox{{\tiny\cite{DR20}}}}$, to avoid confusion) in the present setting are given by  $ \alpha_{\mbox{{\tiny\cite{DR20}}}} := (k+\alpha)^2 $, \DIFF{$ m_{\mbox{{\tiny\cite{DR20}}}} : = |k+\alpha| $}, $ \beta_{\mbox{{\tiny\cite{DR20}}}} : = - \,2\,(k+\alpha)\,S(0) $.
Note however that in \cite{DR18,DR20} the full operator $ H_{\alpha,S} $ is not taken into account and therefore the self-adjoint extensions mixing the subspaces $ k = 0 $ and $ k = -1 $ are not discussed.}
in \cite{DR18,DR20}. In particular, it is proven there that all the operators are essentially self-adjoint, except for the ones with $k = 0 $ and $ k = -1 $, which have both deficiency indices $ (1,1) $ and therefore a one-parameter family of self-adjoint extensions each. 
The defect functions are 
\begin{equation}
	\label{eq: galpha}
g_{\alpha,\pm}^{(k)}(r) := N_{\alpha,\pm}^{(k)}\,W_{-\, e^{\pm i \pi/4} (\alpha + k) S(0),\, |\alpha + k|} \big(2\,e^{\mp i \pi/4}\, r\big)\,,
\end{equation}
where $W_{\kappa,\mu}(z)$ is the Whittaker function related to the Tricomi's function $U(a,b;z) = z^{-a}\,{}_{2}F_{0}(a,a-c+1,-1;-1/z)$ via the identity (see \cite[\S\,13.14, Eq. 13.14.3]{NIST})
\begin{equation}
W_{\kappa,\mu}(z) = e^{-z/2}\,z^{\mu + 1/2}\, U\Big(\, \tx\frac{1}{2} + \mu - \kappa, 1+ 2\mu;z\Big)\,,
\end{equation}
and we have fixed the normalization factor as
\begin{equation}
	\label{eq:Nchoice}
N_{\alpha,\pm}^{(k)} := {\pm\,(e^{\pm i \pi/4})^{1/2-|\alpha+k|} \over \Gamma\left(1/2+|\alpha+k|+ e^{\mp i \pi/4} (\alpha+k)\, S(0) \right)}\,,
\end{equation}
in order to simplify the labelling of the distinguished extensions (see next \cref{rem: friedrichs}).

We are now in position to state the main result of the section. 
We introduce the defect functions
\begin{equation}\label{Upsilondef}
	\Upsilon_{\pm}^{(0)}(r) := \tx{1 \over \sqrt{2\pi}}\;r^{-1/2}\,g_{\alpha,\pm}^{(0)}(r)\,, \qquad
\Upsilon_{\pm}^{(-1)}(r,\theta) := \tx{1 \over \sqrt{2\pi}}\;r^{-1/2}\,g_{\alpha,\pm}^{(-1)}(r)\,e^{-i \theta}\,.
\end{equation}

	\begin{proposition}[Self-adjoint extensions $H^{(U)}_{\alpha,S}$]
		\label{prop: special}
		\mbox{}	\\
		Let $\alpha \in (0,1) $ and let $ \mathbf{S} $ satisfy \eqref{Shyp} and \eqref{Shyp 3}. Then, the symmetric operator $ \dot{H}_{\alpha,S} $ on $ L^2(\R^2) $ admits a four-parameter family of self-adjoint extensions $ H^{(U)}_{\alpha,S} $, $ U \in M_2(\mathbb{C}) $ unitary, given by
  		\bml{
  			\dom \big(H^{(U)}_{\alpha,S}\big) = \lf\{ \psi = \phi \!+\! \Upsilon_{+}\! +\!  \Upsilon_{-} \!\in\! L^2(\mathbb{R}^2) \,\Big|\, \phi \!\in\! \dom\big(\dot{H}_{\alpha,S}\big),\, \Upsilon_{\pm} = c_{0,\pm} \Upsilon_{\pm}^{(0)}\! + c_{-1,\pm}\Upsilon_{\pm}^{(-1)}\;, \ri. \\
  			\lf. \mathbf{c}_{\pm} : = \lf( c_{0, \pm}, c_{-1, \pm} \ri) \!\in\! \mathbb{C}^2,  \mathbf{c}_- = U\, \mathbf{c}_+ \ri\}\,, 
		}
		\beq
			H^{(U)}_{\alpha,S}\, \psi = \dot{H}_{\alpha,S}\,\phi + i\,\Upsilon_{+} - i\,\Upsilon_{-}\,.
		\eeq
	\end{proposition}

    \begin{remark}[Classification of self-adjoint extensions]
    		\label{rem: all sa extensions}
    		\mbox{}	\\
    		As in the usual application of Von Neumann theory, the family $ H^{(U)}_{\alpha,S} $ covers all possible self-adjoint extensions of the operator $ \dot{H}_{\alpha,S} $. Being the deficiency indices equal to $ (2,2) $, the family is labelled by 4 real parameters identifying the entries of the matrix $ U $, which has the form
    		\beq
    			U  = 
    			e^{i\eta} \left(\begin{matrix} a & -\overline{b} \\ b & \overline{a}\end{matrix}\right),
    			\qquad		\mbox{where } \eta \in [0,2\pi), a, b \in \mathbb{C}, \mbox{ s.t. }|a|^2 + | b|^2 = 1.
		\eeq
		The extensions studied in \cite{DR20} corresponds to diagonal matrices  $ U $.
    \end{remark}
    
    \begin{remark}[Friedrichs and Krein extensions]
    		\label{rem: friedrichs}
    		\mbox{}	\\
    		The Friedrichs and Krein extensions obviously belong to the family $ H^{(U)}_{\alpha,S} $ and are recovered for $ U $ equal to
    		\bdm
    			U^{(F)} := \left(\begin{matrix} 1 & 0 \\ 0 & 1\end{matrix}\right)\,, \qquad
    			U^{(K)} := \left(\begin{matrix} -1 & 0 \\ 0 & -1\end{matrix}\right)\,,
    		\edm
		respectively.
    \end{remark}
    
    \begin{remark}[Spectral properties]
    		\label{rem: spectral}
    		\mbox{}	\\
    		The spectral properties of the Whittaker operators $L_{\alpha,S}^{(k)} $  in \eqref{eq: Lalpha} are discussed in detail in \cite[\textsection 3]{DR20}: each operator has the same essential spectrum of the Friedrichs extension, {\it i.e.}, $ \sigma_{\mathrm{ess}}(L_{\alpha,S}^{(k)}) = \R_+ $, for any $ k \in \Z $, and typically isolated eigenvalues appear below the threshold of the essential spectrum. Unfortunately, these results do not directly apply to the operators $ H_{\alpha,S}^{(U)} $ for generic $ \mathbf{S} $, because, as we are going to see (see \cref{sec: special proof}), such operators can be seen as Kato-small perturbations of the corresponding ones with constant $ S(\xv) = S(\mathbf{0}) $ and, as such, their spectral properties depend on $ \mathbf{S} $ in a crucial way.
     \end{remark}
     
     \begin{remark}[Anyonic operators]
     		\label{rem: anyonic}
     		\mbox{}	\\
     		Among the extensions $ H_{\alpha,S}^{(U)} $, there is a one-parameter family identifying anyonic operators, {\it i.e.}, self-adjoint realizations of the Hamiltonian of a pair of anyons immersed in an external magnetic field $ \nabla \times \mathbf{S} $. Such extensions are those living in $ L^2_{\mathrm{even}}(\R^2) $, namely those associated with the defect functions $ \Upsilon_{\pm}^{(0)} $ only, {\it i.e.}, $s$-wave perturbations. 
Their explicit expression is obtained by taking $a = e^{i\tau}$, $b = 0$ and $\eta = \tau$:
\begin{equation*}
U = \left(\begin{matrix} e^{2i\tau} & 0 \\ 0 & 1 \end{matrix}\right).
\end{equation*}
Note that this implies that the domain of such extensions is given by wave functions $ \psi $ such that
\beq
	\label{eq: decomposition}
	\psi = \phi + c_{0} \Big(\Upsilon_{+}^{(0)}\! + e^{2i\tau}\, \Upsilon_{-}^{(0)}\Big)
+ c_{-1}\Big(\Upsilon_{+}^{(-1)}\! + \Upsilon_{-}^{(-1)}\Big),	\qquad		\mbox{with } \phi \in \dom\big(\dot{H}_{\alpha,S}\big),\; c_0, c_{-1} \in \mathbb{C}.
\eeq
It is interesting to remark that the last term in the expansion of $ \psi $ belongs to the domain of the unperturbed operator and can thus be included in $ \phi $. Furthermore, since $ \Upsilon_{\pm}^{(0)} $ have the same asymptotic behavior as $ r \to 0^+ $ of $ G_{\la} $, {\it i.e.}, $ \Upsilon_{\pm}^{(0)}(r) \sim r^{-\alpha} $, any functions as in \eqref{eq: decomposition} above admits a decomposition as in \eqref{domHbe1}, namely $ \psi = \phi_{\la} + q \chi G_{\la} $. Then, it is not difficult to verify that the ``charge'' $ q \in \mathbb{C} $ satisfies the boundary condition in \eqref{domHbe1} and one recovers the operators given in \cref{cor:domHb1} (for suitable $ \beta \in \R $).
	\end{remark}

\section{Proofs}

We now present the proofs of our main results. We start with the investigation of the Friedrichs extension and then study its singular perturbations.

\subsection{Regular realizations: the Friedrichs extension} \label{sec: friedrichs proof}
Before dealing with the presence of an external magnetic field, let us first consider the quadratic form $Q_{\alpha,0}^{(F)}$ in absence of it, {\it i.e.}, for $ \mathbf{S} = \mathbf{0} $. Closedness and non-negativity are obvious consequences of Eqs.\,\eqref{QalS} and \eqref{QFr}. We recall the decomposition in cylindrical harmonics in \eqref{RepSph} (with analogous expressions for $ L^2_{\mathrm{even}}(\R^2) $), which allows to rewrite the form for any $\psi \in \dom[Q_{\alpha,0}^{(F)}]$ as
\beq
Q_{\alpha,0}[\psi] 
= \sum_{k \in \mathbb{Z}} \int_{0}^{+\infty}\!\! \diff r\;r \left[\,|\psi'_{k}|^2\!+ {(k + \alpha)^2 \over r^2}\;|\psi_k|^2\right]; 
\eeq
\beq
\|\nabla \psi\|_{2}^{2}
= \sum_{k \in \mathbb{Z}} \int_{0}^{+\infty}\!\! \diff r\;r \left[\,|\psi'_{k}|^2\!+ {k^2 \over r^2}\;|\psi_k|^2\right], 
\qquad
\| \mathbf{A}_{\alpha} \psi\|_{2}^{2} 
= \sum_{k \in \mathbb{Z}} \int_{0}^{+\infty}\!\! \diff r\;{\alpha^2 \over r}\;|\psi_k|^2\,.
\eeq
Then, the basic inequalities $(k + \alpha)^2 \leqslant 2\,(k^2 + \alpha^2)$ and $(k + \alpha)^2 \geqslant (1-\alpha)^2\,k^2 $ immediately implies that $ (k + \alpha)^2 \geqslant \min\{\alpha^2,(1-\alpha)^2\} $, so that
\beqn
	Q_{\alpha,0}[\psi] & \leqslant & 2\, \lf\|\nabla \psi \ri\|_{2}^{2} + 2 \lf\| \mathbf{A}_{\alpha} \psi \ri\|_{2}^{2}, 	\label{eq: Q0 upper} \\
	Q_{\alpha,0}[\psi] & \geqslant & \epsilon (1-\alpha)^2 \,\lf\|\nabla \psi \ri\|_{2}^{2} + (1 - \epsilon) \min\, \{1,(1-\alpha)^2/\alpha^2\}\,\lf\| \mathbf{A}_{\alpha} \psi \ri\|_{2}^{2}. \label{eq: Q0 lower}
\eeqn
for any $ \epsilon \in [0,1] $. These facts prove that $\dom[Q_{\alpha,0}^{(F)}] = \{\psi \!\in\! H^1(\mathbb{R}^2)\,|\, \mathbf{A}_{\alpha} \psi \!\in\! L^2(\mathbb{R}^2)\}$\,, which is equivalent to Eq.\,\eqref{domQF} for $\mathbf{S} = \mathbf{0}$. On the other hand,  the square-integrability of $ \mathbf{A}_{\alpha}\psi$ and $\nabla \psi$ in a neighborhood of the origin requires that
\beq
	\label{eq: limits at the origin}
\sum_{k \in \mathbb{Z}} |\psi_k(r)|^2 
= r^{-1} \int_{\partial B_{r}}\!\! \diff\Sigma_r\;|\psi|^2\;
\xrightarrow[r \to 0^+]{} \,0\,, 
\qquad
r^2 \sum_{k \in \mathbb{Z}} |\psi'_k(r)|^2 
= r \int_{\partial B_{r}}\!\! \diff\Sigma_r\;|\partial_{r} \psi|^2\;
\xrightarrow[r \to 0^+]{} \,0\,,
\eeq
respectively, which proves Eq.\,\eqref{limdomQF} for any $\psi \in \dom (Q_{\alpha,0}^{(F)})$.

\begin{proof}[Proof of \cref{prop:QF}] \mbox{}	

{\sl i)} 
Closedness and non-negativity of the quadratic form are evident. Let us prove \eqref{domQF}, by showing the reciprocal inclusion of the sets on its l.h.s. and r.h.s.. On one hand, acting exactly as in the derivation of \eqref{eq: Q0 upper}, we get
$$
Q_{\alpha,S}[\psi] \leqslant 2\,\lf\| \lf(- i \nabla + \mathbf{S} \ri)\psi \ri\|_{2}^{2} + 2\, \lf\|\mathbf{A}_{\alpha} \psi \ri\|_{2}^{2}\,,
$$
which suffices to infer that the r.h.s. of \eqref{domQF} is contained in $ \dom[Q_{\alpha,S}^{(F)}] $. 
On the other hand, consider the partition of unity given by a pair of $C^{\infty}$ functions $ \xi,\eta : \mathbb{R}^2 \to [0,1]$ such that $\xi$ has compact support, $\xi = 1$ in an open neighborhood of the origin, and $\xi^2 + \eta^2 = 1$. Then, starting again from \eqref{QalS} and using a variant of the IMS localization formula (see, e.g., \cite[Thm.\,3.2]{Cycon}) and the lower bounds derived previously in the case $ \mathbf{S} = \mathbf{0} $, we derive the following chain of inequalities for any $ \epsilon \in (0,1)$:
\bmln{
	Q_{\alpha,S}[\psi] = \lf\| \lf(-i \nabla \!+\! \mathbf{A}_{\alpha} \!+\! \mathbf{S} \ri)(\xi \psi) \ri\|_{2}^{2} +  \lf\| \lf(-i \nabla \!+\! \mathbf{A}_{\alpha} \!+\! \mathbf{S} \ri)(\eta \psi) \ri\|_{2}^{2} - \lf\| \lf(\nabla \xi \ri)\psi \ri\|_{2}^{2}  - \lf\| \lf(\nabla \eta\ri) \psi \ri\|_{2}^{2} \\
\geqslant (1- \epsilon) \left( \lf\| \lf(-i \nabla \!+\! \mathbf{S} \ri)(\eta \psi) \ri\|_{2}^{2} + \lf\| \lf(-i \nabla \!+\! \mathbf{A}_{\alpha} \ri)(\xi \psi) \ri\|_{2}^{2} \right)  - \tx\frac{1 - \epsilon}{\epsilon} \left( \lf\| \mathbf{S} \chi\, \psi \ri\|_{2}^{2} + \lf\|\mathbf{A}_{\alpha}\eta\, \psi \ri\|_{2}^{2} \right)  - C \lf\| \psi \ri\|_{2}^{2}  \\
\geqslant (1- \epsilon) \left( \lf\| \lf(-i \nabla \!+\! \mathbf{S} \ri)(\eta \psi) \ri\|_{2}^{2} + \epsilon (1-\alpha)^2 \lf\| \nabla (\xi \psi) \ri\|_{2}^{2} + (1-\epsilon) \min\, \{1,(1\!-\!\alpha)^2\!/\alpha^2\}\, \lf\| \mathbf{A}_{\alpha} \xi\, \psi \ri\|_{2}^{2} \right)  \\
- \lf[ \tx\frac{1 - \epsilon}{\epsilon} \left( \lf\| \mathbf{S} \chi \ri\|_{\infty}^2 + \lf\| \mathbf{A}_{\alpha}\eta \ri\|_{\infty}^2  \right)  + C \ri] \lf\| \psi \ri\|_{2}^{2} \\
\geqslant \epsilon(1-\epsilon)^2 (1\!-\!\alpha)^2\! \left( \lf\| \lf(-i \nabla \!+\! \mathbf{S} \ri)(\eta \psi) \ri\|_{2}^{2} + \lf\| \lf(-i \nabla \!+\! \mathbf{S} \ri)(\xi \psi) \ri\|_{2}^{2} \right)\!
+ (1-\epsilon)^2 \min\, \{1,(1\!-\!\alpha)^2\!/\alpha^2\}\, \lf\| \mathbf{A}_{\alpha} \, \psi \ri\|_{2}^{2} \\
 - \lf[ (1\!-\!\epsilon)^2 (1\!-\!\alpha)^2  \lf\| \mathbf{S} \xi \ri\|_{\infty}^2 \! + (1- \epsilon)^2 \min\, \{1,(1\!-\!\alpha)^2\!/\alpha^2\}\, \lf\| \mathbf{A}_{\alpha} \eta\ri\|^2_{\infty} + C_{\epsilon} \ri] \|\psi\|_{2}^{2} \\
\geqslant \epsilon(1-\epsilon)^2 (1\!-\!\alpha)^2\! \lf\| \lf(-i \nabla \!+\! \mathbf{S} \ri) \psi \ri\|_{2}^{2} 
+ (1-\epsilon)^2 \min\, \{1,(1\!-\!\alpha)^2\!/\alpha^2\}\, \lf\| \mathbf{A}_{\alpha} \, \psi \ri\|_{2}^{2} - C_{\epsilon} \lf\|\psi \ri\|_{2}^{2},
}
where $ C_\epsilon > 0 $ is finite for any $ \epsilon \in (0,1) $ and we have used the inequality 
\begin{equation}
	\label{eq: inequality}
|\uv+\vv|^2 
\geqslant (1- \varepsilon)\, |\uv|^2 + \left(1 - \tx{1 \over \epsilon}\right) |\vv|^2,
\end{equation}
following from $|\uv+\vv| \geqslant \big||\uv| - |\vv|\big|$ and $|\uv \cdot \vv| \leqslant {\epsilon \over 2}\,|u|^2 + {1 \over 2\epsilon}\,|\vv|^2$.
Therefore, for all $\gamma > 0$ large enough, there exists a positive constant $C_{\gamma}$ such that
\begin{equation}
Q_{\alpha,S}[\psi] + \gamma\,\|\psi\|_{2}^{2} \geqslant C_{\gamma}\left( \lf\| \lf(-i \nabla \!+\! \mathbf{S} \ri) \psi \ri\|_{2}^{2}  + \lf\| \mathbf{A}_{\alpha} \, \psi \ri\|_{2}^{2} \right),
\end{equation}
which implies the reversed inclusion of $ \dom(Q_{\alpha,S}^{(F)}) $ in the r.h.s. of \eqref{domQF}, whence the identity.

Finally, notice that $ \mathbf{S} \psi \!\in\! L^{2}_{\mathrm{loc}}(\mathbf{R}^2)$ for any $\psi \!\in\! L^2(\mathbf{R}^2)$, since $ \mathbf{S} \!\in\! L^{\infty}_{\mathrm{loc}}(\mathbf{R}^2)$. On account of Eq.\,\eqref{domQF}, this fact grants that $\nabla\psi$ and $ \mathbf{A}_{\alpha}\psi$ both belong to $L^2_{\mathrm{loc}}(\mathbf{R}^2)$ for any $\psi $ in the form domain. Therefore, the limits in Eq.\,\eqref{limdomQF} can be deduced exactly as in the case $ \mathbf{S} = \mathbf{0}$ in \eqref{eq: limits at the origin} above.

{\sl ii)} Consider the sesquilinear form defined by polarization starting from $Q_{\alpha,S}^{(F)}$, that is
$$
	Q_{\alpha,S}^{(F)}[\psi_{1},\psi_{2}] := \!\int_{\mathbb{R}^2}\!\diff\mathbf{x}\;\overline{\lf(-i \nabla + \mathbf{A}_{\alpha} + \mathbf{S} \ri)\psi_{1}} \cdot \lf(-i \nabla + \mathbf{A}_{\alpha} + \mathbf{S} \ri)\psi_{2}, \qquad \mbox{for } \psi_{1},\psi_{2} \in \dom\big[Q_{\alpha,S}^{(F)} \big]\,,
$$
and notice that the usual definition of the unique operator associated to a closed, semi-bounded quadratic form (see, e.g., \cite[Thm.\,VIII.15]{ReSi1}) in the case under analysis yields
$$
	\dom \big(H_{\alpha,S}^{(F)}\big) = \lf\{ \psi_{2} \!\in\! \dom\big[ Q_{\alpha,S}^{(F)} \big] \;\big|\; \exists\,w \!\in\! L^2(\mathbb{R}^2)\;\,\mbox{s.t.}\; Q_{\alpha,S}^{(F)}[\psi_{1},\psi_{2}] = \braketl{\psi_{1}}{w}_{L^2(\mathbb{R}^2)}\;, \forall\,\psi_{1} \!\in\! \dom\big[Q_{\alpha,S}^{(F)}\big] \ri\}\,,
$$
with $H_{\alpha,S}^{(F)} \psi_{2} := w$, for all $\psi_{2} \!\in\! \dom (H_{\alpha,S}^{(F)})$.
For any $\psi_{1},\psi_{2}\! \in\! \dom(Q_{\alpha,S}^{(F)})$, by integrating by parts, we obtain
\begin{align*}
Q_{\alpha,S}^{(F)}[\psi_{1},\psi_{2}] 
& = \lim_{r \to 0^+} \!\int_{\mathbb{R}^2 \setminus B_r}\! \diff\mathbf{x}\;\overline{\lf(-i \nabla + \mathbf{A}_{\alpha} + \mathbf{S} \ri)\psi_{1}} \cdot \lf(-i \nabla + \mathbf{A}_{\alpha} + \mathbf{S} \ri)\psi_{2}, \\
& = \lim_{r \to 0^+} \!\int_{\mathbb{R}^2 \setminus B_r}\!\diff\mathbf{x}\;\overline{\psi_{1}} \lf(-i \nabla + \mathbf{A}_{\alpha} + \mathbf{S} \ri)^2 \psi_{2},\,
- \lim_{r \to 0^+}\! \int_{\partial B_{r}}  \diff\Sigma_r\;\overline{\psi_{1}}\; \lf( \partial_{r} + i\, \mathbf{S} \!\cdot\! \hat{\mathbf{r}} \ri)\, \psi_{2}\;.
\end{align*}
Concerning the boundary term, assumption \eqref{Shyp} and the relations in Eq.\,\eqref{limdomQF} imply
\bmln{
	\left|\int_{\partial B_{r}}\hspace{-0.3cm}\diff \Sigma_r\;\overline{\psi_{1}}\; \lf( \partial_{r} + i\, \mathbf{S} \!\cdot\! \hat{\mathbf{r}} \ri)\, \psi_{2} \right|	
	\leqslant \left(\!\int_{\partial B_{r}}\hspace{-0.3cm}\diff\Sigma_r\; |\psi_{1}|^2\!\right)^{\!\!\!1/2}\left[
\left(\!\int_{\partial B_{r}}\hspace{-0.3cm}\diff\Sigma_r\; \lf|\partial_r \psi_{2}\ri|^2\!\right)^{\!\!\!1/2}
+ \|\mathbf{S}\|_{L^{\infty}(B_r)}\! \left(\!\int_{\partial B_{r}}\hspace{-0.3cm}\diff\Sigma_r\; |\psi_{2}|^2\!\right)^{\!\!\!1/2}
\right]\\
= 2 \pi \sqrt{\avg{\lf| \psi_1 \ri|^2}(r)} \lf[ r \sqrt{\avg{\lf| \partial_r \psi_2 \ri|^2} (r)} + r \|\mathbf{S}\|_{L^{\infty}(B_r)} \sqrt{\avg{\lf| \psi_2 \ri|^2}(r)} \ri]	 \xrightarrow[r \to 0^+]{} 0\,,
}
where we recall that $ \avg{ f}: \R^+ \to \mathbb{C} $ stands for the angular average of the function $ f: \R^2 \to \mathbb{C} $ (see \eqref{eq: avg}). In view of Eq.\,\eqref{HalS} and \eqref{domQF}, the above results suffice to infer that $H_{\alpha,S}^{(F)} \psi_{1} = H_{\alpha,S} \psi_{1}$, which in turn implies Eq.\,\eqref{domHF}.
\end{proof}

Before proceeding further, we briefly comment on what is stated in \cref{rem: bounded S}. The first identity in Eq.\,\eqref{domHF0} is a trivial consequence of Eq.\,\eqref{domQF} and of the fact that $ \mathbf{S} \psi \in L^2(\mathbb{R}^2)$ for any $ \mathbf{S} \in L^{\infty}(\mathbb{R}^2)$ and $\psi \in L^2(\mathbb{R}^2)$. Concerning the second identity in Eq.\,\eqref{domHF0}, the inclusion $ \dom(H_{\alpha,0}^{(F)}) \subset\dom (H_{\alpha,S}^{(F)})$ can be easily deduced taking into account that $ \dom (H_{\alpha,S}^{(F)}) \subset \dom (Q_{\alpha,S}^{(F)}) $ and noting that
\begin{align*}
	\lf\|H_{\alpha,S} \psi \ri\|_{2}^{2} 
& \leqslant 3 \int_{\mathbb{R}^2}\!\! \diff \mathbf{x} \left[\,\lf| \lf(- i \nabla + \mathbf{A}_{\alpha} \ri)^{2}\psi \ri|^2 + 4 \,\lf| \mathbf{S} \!\cdot\! \lf(-i \nabla + \mathbf{A}_{\alpha} \ri)\psi \ri|^2 + \lf| \mathbf{S}^2\, \psi \ri|^2\right] \\
& \leqslant 3 \,\lf( \lf\|H_{\alpha,0} \psi \ri\|_{2}^{2} + 4 \,\lf\| \mathbf{S} \ri\|^2_{\infty}\, Q_{\alpha,0}[\psi] + \lf\|\mathbf{S}\ri\|^4_{\infty} \|\psi\|_{2}^{2} \ri)\,.
\end{align*}
On the other hand, by \eqref{eq: inequality},
\begin{align*}
\|H_{\alpha,S} \psi\|_{2}^{2} 
& \geqslant \int_{\mathbb{R}^2}\!\!\diff \mathbf{x}\left[(1-\epsilon)\,\lf| \lf(- i \nabla + \mathbf{A}_{\alpha} \ri)^{2}\psi \ri|^2 + \left(1- \tx{1 \over \epsilon}\right) \lf| \lf( 2 \mathbf{S} \!\cdot\! \lf(-i \nabla + \mathbf{A}_{\alpha} \ri)  + \mathbf{S}^2 \ri)\, \psi \ri|^2 \right] \\
& \geqslant (1-\varepsilon)\,\|H_{\alpha,0}\psi\|_{2}^{2} + 3 \left(1- \tx{1 \over \epsilon}\right) \left( 2\,\lf\| \mathbf{S} \ri\|^2_{\infty}\, Q_{\alpha,0}[\psi] + \lf\|\mathbf{S}\ri\|^4_{\infty} \|\psi\|_{2}^{2} \right),
\end{align*}
for any $ \epsilon \in (0,1) $, which implies the reversed inclusion $ \dom (H_{\alpha,S}^{(F)}) \subset \dom (H_{\alpha,0}^{(F)})$.

\subsection{Singular perturbations ($ \mathbf{S} $ continuous)}\label{sec: singular proof}
We first discuss in detail the well-posedness of the quadratic form \eqref{Qbeta}, which we recall is given by
\bmln{
Q^{(\beta)}_{\alpha,S}[\psi] 
:= Q_{\alpha,S}^{(F)}[\phi_{\lambda}] -\,\lambda^2\,\|\psi\|_{2}^{2} + \lambda^2\, \|\phi_{\lambda}\|_{2}^{2} \\
 + 2 \,\Re \lf[q\,\lf(
2\,  \braketr{ \lf(-i \nabla \!+\! \mathbf{A}_{\alpha} \ri)\,\phi_{\lambda}}{\lf(\mathbf{S}\,\chi - i\nabla \chi\ri)\,G_{\lambda}}
+ \braketr{ \phi_{\lambda}}{\lf(\mathbf{S}^2 \chi + \Delta \chi \ri) G_{\lambda}} \ri) \ri]  \\
 + |q|^2\, \lf(\beta + c_{\alpha}\, \lambda^{2 \alpha} + \Xi_{\alpha,S}(\lambda) \ri)\;, 
} 
	\bdm
	\Xi_{\alpha,S}(\lambda) := \braketr{ G_{\lambda}}{\lf(\mathbf{S}^2 \chi^2 -\chi\,\Delta \chi \ri) G_{\lambda} }
	- \braketr{ G_{\lambda}}{\nabla \chi^2 \cdot \nabla G_{\lambda}}
	+ 2\, \braketr{ G_{\lambda}}{ \mathbf{S}\!\cdot\!\mathbf{A}_{\alpha}\,\chi^2 G_{\lambda}}\,, 
\edm
so completing what is stated in \cref{rem:L2inner}. Besides the first term on the second line of \eqref{Qbeta}, there are indeed other terms in the form, which may seem to be ill-defined. 

First of all, we observe that, since $ \mathbf{S}^2 \,\chi\,, \Delta \chi \!\in\! L^{\infty}(\mathbb{R}^2)$ by construction, the expressions $ \braketr{\phi_{\lambda}}{(\mathbf{S}^2\, \chi + \Delta \chi) G_{\lambda}}$ and $ \braketr{\chi G_{\lambda}}{(\mathbf{S}^2 \,\chi -\Delta \chi) G_{\lambda}} $ are actually well-defined inner products in $L^2(\mathbb{R}^2)$. Next, we remark that our assumptions on $\chi$ imply $\nabla \chi^2 \equiv 0$ in an open neighborhood of the origin. On the other hand, it can be checked by direct computation that $\nabla G_{\lambda} \!\in\! L^2_{\mathrm{loc}}(\mathbb{R}^2 \setminus \{\mathbf{0}\})$. Hence, we deduce that $\nabla \chi^2 \cdot \nabla G_{\lambda} \in L^2(\mathbb{R}^2)$, so that the second term in $ \Xi_{\alpha,S}(\lambda)$ is well-posed.

In order to show that, under our assumptions, $ \Xi_{\alpha,S}(\lambda) $ is finite and thus the form is well-posed, it remains to consider the last term in its expression and, in light of \cref{rem: regularity}, we should expect that it is the most delicate one. Recalling the asymptotics \eqref{eq: asympt 0} and, specifically, the fact that  $| \mathbf{A}_{\alpha}| \sim r^{-1}$ and $G_{\lambda} \sim r^{-\alpha}$, as $r \to 0^{+}$, we see, that for $0 \!<\! \alpha \!<\! 1/2$ and for any $ \mathbf{S} \!\in\! L^{\infty}_{\mathrm{loc}}(\mathbb{R}^2)$, such an expression can be interpreted as a duality pairing between the weighted spaces $L^2(\mathbb{R}^2,|\mathbf{x}|^{-1}\diff\mathbf{x})$ and $L^2(\mathbb{R}^2,|\mathbf{x}|\,\diff\mathbf{x})$. 
When $1/2 \!\leqslant\! \alpha \!<\! 1$, on the other hand, we need stronger assumptions on the regularity of $ \mathbf{S} $ at the origin. However, H\"older continuity, as in \eqref{Shyp 2}, allows to give a meaning to that expression as a duality pairing between  $L^2(\mathbb{R}^2,\,|\mathbf{x}|^{\,-\,(1-\nu)}\diff\mathbf{x})$ and $L^2(\mathbb{R}^2,\,|\mathbf{x}|^{1-\nu} \diff\mathbf{x})$.
It is worth noting that the H\"older condition on $\mathbf{S}$ is not an optimal requirement in general, since the last term in $ \Xi_{\alpha,S}(\lambda) $ totally disappears for, {\it e.g.}, radial magnetic potentials, {\it i.e.}, whenever $ \mathbf{S}(\xv) = S(\xv) \,\hat{\xv} $.

\begin{proof}[Proof of \cref{prop:Qbeta}]	\mbox{}

{\sl i)} Let us first prove the independence of the form of $ \la > 0 $. Let us fix $\lambda_1 \neq \lambda_2$ and consider, for any $\psi \in \dom\big(Q^{(\beta)}_{\alpha,S}\big)$, the two alternative representations $\psi = \phi_{\lambda_1} \!+ q\,\chi\,G_{\lambda_1}$ and $\psi = \phi_{\lambda_2} \!+ q\,\chi\,G_{\lambda_2}$, which follow from the fact that $G_{\lambda_2}\! - G_{\lambda_1} \!\in\! \dom(Q_{\alpha,S}^{(F)})$, so that $\phi_{\lambda_1} \!= \phi_{\lambda_2} \!+ q\,\chi\,(G_{\lambda_2}\! - G_{\lambda_1})$. Note that a non-trivial consequence is that the ``charge'' is independent of $ \lambda $. Moreover, the identity \eqref{greeeq} and some integrations by parts yield
\bml{
	Q^{(\beta)}_{\alpha,S}\big[\phi_{\lambda_1} \!+\! q\,\chi\,G_{\lambda_1}\big]
	= Q_{\alpha,S}^{(F)}\big[\phi_{\lambda_2}\big] 
- \lambda_2^2\,\|\psi\|_{2}^{2} + \lambda_2^2\, \lf\|\phi_{\lambda_2} \ri\|_{2}^{2}  \\
	+ 2 \,\Re \lf[q\,\lf(2\, \braketr{ \lf(-i \nabla \!+\! \mathbf{A}_{\alpha} \ri)\,\phi_{\lambda_2}}{\, \lf(\mathbf{S}\,\chi - i\nabla \chi \ri)\,G_{\lambda_2}} + \braketr{\phi_{\lambda_2}}{\lf(\mathbf{S}^2\,\chi + \Delta \chi \ri) G_{\lambda_2}} \ri) \ri]  \\
	+ |q|^2\,\left[\beta + c_{\alpha}\, \lambda_2^{2\alpha} + \Xi_{\alpha,S}(\lambda_2) 
	-i \int_{\mathbb{R}^2}\!\diff\mathbf{x}\;\lf( \mathbf{A}_{\alpha}\!\cdot\! \nabla \chi^2 \ri) \lf(G_{\lambda_2} - G_{\lambda_1} \ri)^{2} 
	+ \,c_{\alpha}\, (\lambda_1^{2\alpha}\! - \lambda_2^{2\alpha}) \right.  \\
	\left. + (\lambda_2^2 - \lambda_1^2)\, \braketr{\chi G_{\lambda_2}}{\chi G_{\lambda_1}} + \!\int_{\mathbb{R}^2}\!\!\diff\mathbf{x}\, \lf(\nabla \chi^2\ri) \!\cdot\! \lf(G_{\lambda_1}\! \nabla G_{\lambda_2}\! - G_{\lambda_2}\!\nabla G_{\lambda_1} \ri)
\right] \\
	- 2\, \Re \!\left[q\, \lim_{r \to 0^+} \int_{\partial B_{r}}\!\!\! \diff\Sigma_{r}\; \overline{\phi_{\lambda_2}\!}\; \lf(\chi\,\partial_r - \partial_r \chi - i\, \lf(\mathbf{S}\cdot \hat{\mathbf{r}} \ri)\;\chi \ri) \lf(G_{\lambda_2}\! - G_{\lambda_1}\ri) \right] 	 \\
	- |q|^2 \!\lim_{r \to 0^+} \int_{\partial B_{r}}\!\!\! \diff\Sigma_{r}\; \chi\,\lf(G_{\lambda_2}\! - G_{\lambda_1}\ri)\,\lf[4\,(\partial_r \chi)\,G_{\lambda_1} + \partial_r \lf(\chi(G_{\lambda_2}-G_{\lambda_1}) \ri) \ri]. \label{l1l2}
}

Notice that $\mathbf{A}_{\alpha} \cdot \hat{\mathbf{r}} = 0$ and, since $G_{\lambda}$ is radial, $\mathbf{A}_{\alpha} \cdot \nabla \lf(G_{\lambda_2} - G_{\lambda_1} \ri)^{2} = 0 $. On account of these identities, recalling as well that $\nabla \cdot \mathbf{A}_{\alpha} = 0$ and integrating by parts we get
\bml{
	\label{eq: ind proof 1}
	\int_{\mathbb{R}^2}\!\diff\mathbf{x} \lf( \mathbf{A}_{\alpha}\!\cdot\! \nabla \chi^2 \ri) \lf(G_{\lambda_2} - G_{\lambda_1} \ri)^{2} = \lim_{r \to 0^+}  \int_{\mathbb{R}^2 \setminus B_{r}}\!\!\!\diff\mathbf{x} \lf( \mathbf{A}_{\alpha}\!\cdot\! \nabla \chi^2 \ri) \lf(G_{\lambda_2} - G_{\lambda_1} \ri)^{2} \\
	= \lim_{r \to 0^+} \lf[ - \int_{\partial B_{r}}\!\! \diff\Sigma_{r}\, \lf( \mathbf{A}_{\alpha} \cdot \hat{\mathbf{r}} \ri)\,\chi^2 \lf(G_{\lambda_2} - G_{\lambda_1} \ri)^{2} - \int_{\mathbb{R}^2 \setminus B_{r}}\!\!\!\diff\mathbf{x}\;  \chi^2 \mathbf{A}_{\alpha} \cdot \nabla\lf(G_{\lambda_2} - G_{\lambda_1} \ri)^{2} \ri] = 0\,.
}
Moreover Eqs.\,\eqref{greeeq} and \eqref{asy0} ensue, respectively, $\Delta G_{\lambda} = (\mathbf{A}_{\alpha}^2 + \lambda^2) G_{\lambda}$ in $\mathbb{R}^2 \setminus \{\mathbf{0}\}$ and (see also \cite[Eq.\,5.5.3]{NIST})
\bdm
G_{\lambda_1}(r)\, \partial_r G_{\lambda_2}(r) - G_{\lambda_2}(r)\, \partial_r G_{\lambda_1}(r)
= {\pi\,(\lambda_1^{2 \alpha} - \lambda_2^{2 \alpha}) \over 2\,\sin(\pi\,\alpha)\,r} + \OO(r^{1-2\alpha}), 
\qquad  \mbox{as } r \to 0^+\,.
\edm
Exploiting these relations and integrating by parts (recall that $c_{\alpha} = \pi^2/\sin(\pi\alpha)$), we obtain
\bml{
	\label{eq: ind proof 2}
c_{\alpha}\, (\lambda_1^{2\alpha}\! - \lambda_2^{2\alpha}) + (\lambda_2^2 - \lambda_1^2)\, \braketr{\chi G_{\lambda_2}}{\,\chi G_{\lambda_1}} + \!\int_{\mathbb{R}^2}\!\!\!\diff\mathbf{x}\, \lf(\nabla \chi^2\ri) \!\cdot\! \lf(G_{\lambda_1}\! \nabla G_{\lambda_2}\! - G_{\lambda_2}\!\nabla G_{\lambda_1} \ri) \\
	= c_{\alpha}\, (\lambda_1^{2\alpha} - \lambda_2^{2\alpha}) - \lim_{r \to 0^+} \int_{\partial B_{r}}\!\!\! \diff \Sigma_{r}\;\chi^2 \lf( G_{\lambda_1}\, \partial_r G_{\lambda_2} - G_{\lambda_2}\, \partial_r G_{\lambda_1} \ri) \\
	+ \lim_{r \to 0^+} \int_{\mathbb{R}^2 \setminus B_r}\!\!\!\diff\mathbf{x}\;\chi^2\, \lf[(\lambda_2^2 - \lambda_1^2)\, G_{\lambda_2}G_{\lambda_1} - G_{\lambda_1} \Delta G_{\lambda_2} + G_{\lambda_2} \Delta G_{\lambda_1} \ri] \\
	= {\pi^2 \over \sin(\pi\alpha)}\, (\lambda_1^{2\alpha} - \lambda_2^{2\alpha}) - \lim_{r \to 0^+} \lf [2 \pi\, r\; \lf({\pi\,(\lambda_1^{2 \alpha} - \lambda_2^{2 \alpha}) \over 2\,\sin(\pi\,\alpha)\,r} + \OO(r^{1-2\alpha}) \ri) \ri] 
= 0\,.
}
Finally, consider the boundary terms in Eq.\,\eqref{l1l2}. Notice that all the derivatives of $\chi$ vanish identically in an open neighborhood of the origin. Then, on account of the first relation written in Eq.\,\eqref{limdomQF} and of the asymptotic expansion \eqref{asy0} for $G_{\lambda}$, in the limit for $r \to 0^+$, we obtain
\bml{
	\label{eq: ind proof 3}
	\left| \int_{\partial B_{r}}\!\!\! \diff\Sigma_{r}\; \overline{\phi_{\lambda_2}\!}\; \lf(\chi\,\partial_r - \partial_r \chi - i\, \lf( \mathbf{S} \cdot \hat{\mathbf{r}} \ri)\;\chi \ri) (G_{\lambda_2}\! - G_{\lambda_1}) \right| 	
	\leqslant 2\pi r \sqrt{\avg{\lf| \phi_{\la_2} \ri|^2}}
\left[\sqrt{\avg{\lf| \partial_r(G_{\lambda_2}\! - G_{\lambda_1}) \ri|^2}}  \ri.	\\
	\lf. + \|\mathbf{S}\,\chi\|_{\infty}\! \sqrt{\avg{\lf| (G_{\lambda_2}\! - G_{\lambda_1}) \ri|^2}}  \right]  
	\leqslant C\, \sqrt{\avg{\lf| \phi_{\la_2} \ri|^2}}
\left[r^{\alpha} + \|\mathbf{S}\,\chi\|_{\infty}\, r^{1+\alpha}\right] \xrightarrow[r \to 0^+]{} 0\,; 
}
\beq
	\label{eq: ind proof 4}
	\int_{\partial B_{r}}\!\!\! \diff\Sigma_{r}\; \chi\,\lf(G_{\lambda_2}\! - G_{\lambda_1}\ri)\,\lf[4\,(\partial_r \chi)\,G_{\lambda_1} + \partial_r \lf(\chi(G_{\lambda_2}-G_{\lambda_1}) \ri) \ri] = \pi\,r  \partial_r(G_{\lambda_2}\! - G_{\lambda_1})^2  = \OO(r^{2\alpha})  \xrightarrow[r \to 0^+]{} 0\,.
\eeq
Summing up \eqref{eq: ind proof 1} -- \eqref{eq: ind proof 4} with \eqref{l1l2}, we obtain that $Q^{(\beta)}_{\alpha,S}\big[\phi_{\lambda_1} \!+\! q\,\chi G_{\lambda_1}\big] = Q^{(\beta)}_{\alpha,S}\big[\phi_{\lambda_2} \!+\! q\,\chi G_{\lambda_2}\big]$, which proves the thesis. 

In the very same way one proves the independence of the cut-off $ \chi $, provided the conditions \eqref{eq: chi} are satisfied. We omit the details for the sake of brevity.

{\sl ii)} To begin with, let us show that $Q^{(\beta)}_{\alpha,S}$ is bounded from below.
In the sequel we denote by $ \one_{\chi}$ the characteristic function of $\mbox{supp}\,\chi$, namely, $ \one_{\chi} : = \one_{\supp \chi} $, and indicate by $C$ a suitable positive constant, which may vary from line to line. Furthermore, all the relations derived in the sequel are understood to be fulfilled for $\lambda > 0$ large enough.

Firstly, exploiting \eqref{eq: inequality}, for any $ \epsilon \in (0,1) $, we get
\bml{
	\label{eq: boundedness proof 1}
Q_{\alpha,S}^{(F)}[\phi_{\lambda}] 
	\geqslant \tx{1 \over 2}\, Q_{\alpha,S}^{(F)}[\phi_{\lambda}] 
+ \tx{1 \over 2}\, \lf \| \one_{\chi} \lf(-i\nabla + \mathbf{A}_{\alpha} + \mathbf{S} \ri)\phi_{\lambda} \ri\|_{2}^{2} \\
	\geqslant \tx{1 \over 2}\, Q_{\alpha,S}^{(F)}[\phi_{\lambda}] 
+ \tx\frac{1 - \epsilon}{2}\, \lf \| \one_{\chi} \lf(-i\nabla + \mathbf{A}_{\alpha}  \ri)\phi_{\lambda} \ri\|_{2}^{2}
- \tx\frac{1 - \epsilon}{2\epsilon}\, \lf\| \one_{\chi} \mathbf{S} \ri\|_{\infty}^2\,\|\phi_{\lambda}\|_{2}^{2}\,.
}
Secondly, we apply the Cauchy inequality, to get, for any $ \epsilon_1 \in (0,1) $,
\bml{
	\label{eq: boundedness proof 2}
 	\Re \lf[ q \, \braketr{ \lf(-i \nabla + \mathbf{A}_{\alpha} \ri)\phi_{\lambda}}{\lf( \mathbf{S}\,\chi - i\nabla \chi \ri)\,G_{\lambda}} \ri] 
	\geqslant -\, \tx{\epsilon_1 \over 2}\, \lf\| \one_{\chi}\, \lf(-i \nabla + \mathbf{A}_{\alpha}\ri) \phi_{\lambda} \ri\|_{2}^{2}
- \tx{1 \over 2\epsilon_1}\,|q|^2\, \lf\| \lf( \mathbf{S} \,\chi - i\nabla \chi \ri)\,G_{\lambda} \ri\|_{2}^{2} \\
	\geqslant -\, \tx{\epsilon_1 \over 2}\, \lf\| \one_{\chi}\, \lf(-i \nabla + \mathbf{A}_{\alpha}\ri) \phi_{\lambda} \ri\|_{2}^{2}
- \tx{1 \over \epsilon_1}\,|q|^2\, \lf( \lf\| \mathbf{S} \,\chi \ri\|_{\infty}^{2} + \lf\|\nabla \chi\ri\|_{\infty}^{2} \ri)\;\|G_{\lambda}\|_{2}^{2} \\
	\geqslant -\, \tx{\epsilon_1 \over 2}\, \lf\| \one_{\chi}\, \lf(-i \nabla + \mathbf{A}_{\alpha}\ri) \phi_{\lambda} \ri\|_{2}^{2}
- - \tx{C \over \epsilon_1}\,|q|^2\, \lambda^{2\alpha-2}\,.
}
Similarly, for any $ \epsilon_2 > 0$,
\bml{
	\label{eq: boundedness proof 3}
	\Re \lf[ q \, \braketr{\phi_{\lambda}}{\lf(\mathbf{S}^2 \chi + \Delta \chi \ri)\,G_{\lambda}} \ri]
	\geqslant -\,\tx{\epsilon_2 \over 2}\,\|\phi_{\lambda}\|_{2}^{2} - \tx{1 \over 2\,\epsilon_2}\,|q|^2\, \lf\| \lf( \mathbf{S}^2 \chi + \Delta \chi \ri)\,G_{\lambda} \ri\|_{2}^{2} \\
	\geqslant  -\,\tx{\epsilon_2 \over 2}\,\|\phi_{\lambda}\|_{2}^{2} - \tx{1 \over \epsilon_2}\,|q|^2\, \lf( \lf\| \mathbf{S}^2 \chi \ri\|_{\infty}^{2} + \lf\|\Delta \chi \ri\|_{\infty}^{2} \ri)\;\|G_{\lambda}\|_{2}^{2} \geqslant - \,\tx{\epsilon_2 \over 2}\,\|\phi_{\lambda}\|_{2}^{2}
- \tx{C \over \epsilon_2}\,|q|^2\, \lambda^{2\alpha-2}\,.
}
Finally, let us consider the term $\Xi_{\alpha,S}(\lambda)$ defined in Eq.\,\eqref{defXi}: since $ \mathbf{S}\, \chi$, as well as $\chi$ and all its derivatives, belongs to $L^{\infty}(\mathbb{R}^2)$, we get
\beq
	\label{eq: boundedness proof 4}
	\Xi_{\alpha,S}(\lambda)  \geqslant - \lf( \lf\| \mathbf{S}\, \chi \ri\|_{\infty}^{2} \!+ \lf\|\Delta \chi \ri\|_{\infty}\ri) \|G_{\lambda}\|_{2}^{2}
- \lf|  \braketr{G_{\lambda}}{\nabla \chi^2 \cdot \nabla G_{\lambda}} \ri|
- 2\,\lf| \braketr{G_{\lambda}}{\,\mathbf{S}\!\cdot\!\mathbf{A}_{\alpha} \,\chi^2 G_{\lambda}} \ri|.
\eeq

Thanks to our assumptions on $\chi$, $\nabla \chi^2 = 0$ inside the disc $B_{r_{1}}$. Hence, in view of the explicit expression \eqref{greenexp} for $G_{\lambda}$ and of the regularity features on the Bessel function $K_{\alpha}$ (see \cite[\S\,10.25]{NIST}), a simple rescaling of the integration variable yields
\bml{
	\label{eq: boundedness proof 5}
	\lf|  \braketr{G_{\lambda}}{\lf(\nabla \chi^2 \ri) \!\cdot\! \nabla G_{\lambda}} \ri|
	\leqslant \lf\|\nabla \chi^2 \ri\|_{\infty}\!\int_{\mathbb{R}^2 \setminus B_{r_{1}}} \diff\mathbf{x}\; \lf|G_{\lambda} \nabla G_{\lambda} \ri|	\leqslant C \lambda^{2\alpha-1}  \int_{\lambda\,r_1}^{+\infty} \diff r\; r\,\lf|K_{\alpha}(r)\, \partial_{r}K_{\alpha}(r) \ri|	\\
\leqslant C\,\lambda^{2\alpha - 1}\,.
}
In order to deal with the last term on the r.h.s. of \eqref{eq: boundedness proof 4}, we have to exploit the regularity of $ \mathbf{S} $: we observe that \eqref{Shyp 2} implies that
\begin{equation}\label{Sleq}
	\lf| \mathbf{S}(\mathbf{x}) \ri| \leqslant C\,|\mathbf{x}|^{\nu}, \qquad \mbox{for all } |\mathbf{x}| \leqslant r_{1}\,,
\end{equation}
where $ \nu > 2 \alpha - 1 $, if $ \alpha \in [1/2,1) $, and $\nu = 0$ otherwise. Hence,
\bml{
	\label{eq: boundedness proof 6}
	\lf| \braketr{G_{\lambda}}{\,\mathbf{S}\!\cdot\!\mathbf{A}_{\alpha} \,\chi^2 G_{\lambda}} \ri|
\leqslant C\, \int_{B_{r_1}}\!\!\! \diff\mathbf{x}\;|\mathbf{x}|^{\nu}\,| \mathbf{A}_{\alpha}|\; G_{\lambda}^2
+ \lf\| \mathbf{S} \chi^2 \ri\|_{\infty}\! \int_{\mathbb{R}^2 \setminus B_{r_1}} \diff \mathbf{x}\;|\mathbf{A}_{\alpha}|\, G_{\lambda}^2 \\
	\leq C \lf[ \lambda^{2\alpha-1-\nu} \int_{0}^{\lambda\,r_{1}}\!\!\! \diff r\;r^{\nu}\, K_{\alpha}^2(r) + \lambda^{2\alpha-1} \int_{\lambda\,r_1}^{+\infty} \diff r\; K_{\alpha}^2(r) \ri]	\leqslant C\,\lambda^{2\alpha-1}\,.
}
Putting together \eqref{eq: boundedness proof 5} and \eqref{eq: boundedness proof 6} with \eqref{eq: boundedness proof 4}, we obtain
\begin{equation}
	\label{eq: xi est proof}
	\Xi_{\alpha,S}(\lambda) \geqslant -\, C\,\lambda^{2\alpha-1}\,.
\end{equation}

For all $\lambda > 0$ sufficiently large, combing \eqref{eq: boundedness proof 1} -- \eqref{eq: boundedness proof 3} with the above \eqref{eq: xi est proof}, we estimate
\bml{
Q^{(\beta)}_{\alpha,S}[\psi] \geqslant \tx{1 \over 2}\, Q_{\alpha,S}^{(F)}[\phi_{\lambda}] -\,\lambda^2\,\|\psi\|_{2}^{2} + \tx{1 \over 2}\,(1 - \epsilon -\,4\epsilon_1)\, 
\lf\| \one_{\chi} \lf(-i\nabla \!+\! \mathbf{A}_{\alpha} \ri)\phi_{\lambda} \ri\|_{2}^{2}  
	\\
	+ \lf(\lambda^2 - \tx{1 - \epsilon \over 2\epsilon}\, \lf\| \one_{\chi} \mathbf{S} \ri\|_{\infty}^2 -\,\epsilon_2 \ri)\, \|\phi_{\lambda}\|_{2}^{2}  + c_{\alpha}\,\lambda^{2 \alpha} \left[ 1 -\,C\,\lf(\lambda^{-1} + \lf(4\,\epsilon_1^{-1} + 2\,\epsilon_2^{-1} \ri)\, \lambda^{-2} \ri) + \tx{\beta \over c_{\alpha}}\,\lambda^{-2 \alpha} \ri] |q|^2\;. \label{Qlower}
}
Now, by the positivity of $Q_{\alpha,S}^{(F)}$, if we take $ \epsilon_1  = \frac{1}{4} (1 - \epsilon) $ and $\lambda > 0$ large enough, so that the coefficients of the last two terms on the r.h.s. are positive, the above relation implies that $Q^{(\beta)}_{\alpha,S}[\psi] \geqslant - \lambda^2\, \|\psi\|_{2}^{2} $, {\it i.e.}, the form is bounded from below. Furthermore, for $ \lambda $ large enough, we also deduce that there exists $ c > 0 $ such that
\begin{align*}
	Q^{(\beta)}_{\alpha,S}[\psi] + \lambda^2\,\|\psi\|_{2}^{2} \geqslant c \,\lf[Q_{\alpha,S}^{(F)}[\phi_{\lambda}] + \lambda^2\, \|\phi_{\lambda}\|_{2}^{2} + \,\lambda^{2\alpha}\,|q|^2 \ri] \,,
\end{align*}
{\it i.e.}, the quadratic form is also coercive. This allows to prove closedness of $Q^{(\beta)}_{\alpha,S}$ retracing the same arguments in \cite[Proof of Thm. 2.4]{CoOd}.

{\sl iii)} If $ \mathbf{S} \in L^{\infty}(\R^2) $ and $ \alpha \in (0,1/2) $, we can start from the expression \eqref{Qbetasmall} of $\widetilde{Q}^{(\beta)}_{\alpha,S}$ and follow the same arguments as in the proof of item {\sl ii)}, to get, for any $\epsilon,\eta \in (0,1)$,
\begin{align}
\widetilde{Q}^{(\beta)}_{\alpha,S}[\psi] 
& \geqslant - \lambda^2\,\|\psi\|_{2}^2 + \| \mathbf{S}\,\psi\|_{2}^2 + (1-\epsilon-2\eta)\,Q_{\alpha,0}^{(F)}[\phi_{\lambda}] + \lf(\lambda^2 - {1 \over \epsilon}\, \lf\| \mathbf{S} \ri\|_{\infty}^2\! \ri)\, \|\phi_{\lambda}\|_{2}^2  \nonumber \\
& \qquad + c_{\alpha}\, \lambda^{2 \alpha}\, \lf(1 - \pi\,\alpha\,\tan(\pi\, \alpha) \,\lf\| \mathbf{S} \ri\|_{\infty}\, \lambda^{-\,1} - {2\alpha\, \over \eta}\,\lf\| \mathbf{S} \ri\|_{\infty}^2 \, \lambda^{-\,2} + {\beta \over c_{\alpha}}\, \lambda^{-2 \alpha} \ri)\,|q|^2 \,,\label{Qlower0}
\end{align}
where we have used the following relation, which can be inferred from \eqref{defXismall} by direct computations:
\bdm
\tilde{\Xi}_{\alpha,S}(\lambda) \geqslant - \,2 \lf\| \mathbf{S} \ri\|_{\infty}\! \int_{\mathbb{R}^2}\!\!\diff\mathbf{x}\, \lf| \mathbf{A}_{\alpha} \ri|\, G_{\lambda}^2 
	= - \,4\pi\,\alpha \lf\| \mathbf{S} \ri\|_{\infty} \lambda^{2\alpha}\! \int_{0}^{\infty}\!\!\!\!\diff r\; K_{\alpha}^2(\lambda\, r) 
= - \,\pi\,\alpha\,\tan(\pi\, \alpha)\,c_{\alpha}\, \lf\| \mathbf{S} \ri\|_{\infty}\, \lambda^{2\alpha-1}\,.
\edm
Here we used the identity (see \cite[Eq.\,6.576.4]{Grad} and recall that we are assuming $\alpha < 1/2$)
\[
\int_{0}^{\infty}\!dr\; K_{\alpha}^2(\lambda\, r) = \tx{1 \over 4\lambda}\,\Gamma^2\!\lf({1 \over 2}\ri)\Gamma\lf({1 \over 2} + \alpha\ri)\Gamma\lf({1 \over 2} - \alpha\ri)\,F\lf({1 \over 2} + \alpha\,,{1 \over 2},1;\,0\ri),
\]
involving the Gauss hypergeometric function $F(a,b,c;z) \equiv {}_{2}F_{1}(a,b,c;z)$ (see \cite[\S 15.1]{NIST}), together with the basic relations $\Gamma(1/2) = \sqrt{\pi}$\,, $\Gamma(1/2+\alpha)\,\Gamma(1/2-\alpha) = \pi/ \cos(\pi\,\alpha)$ and ${}_{2}F_{1}(a,b,c;0) = 1$ (for any choice of the parameters).

Since $Q_{\alpha,0}^{(F)}$ is non-negative, for any $\epsilon,\eta \!>\! 0$, such that $\epsilon + 2\eta \!<\! 1$, and for all $\lambda \geqslant \max\lf\{\lf\| \mathbf{S} \ri\|_{\infty}/\sqrt{\epsilon},\lambda_{*} \ri\}$, with $\lambda_{*} $ implicitly defined by \eqref{lamst}, from \eqref{Qlower0} we deduce
\begin{equation*}
\widetilde{Q}^{(\beta)}_{\alpha,S}[\psi] \geqslant - \,\lambda^2\, \|\psi\|_{2}^2 \,,
\end{equation*}
which yields \eqref{lower1}, in view of the considerations reported in \cref{rem: bounded}.
\end{proof}

Once the quadratic forms are proven to be closed and bounded from below, the derivation of the corresponding self-adjoint operators is done in a standard way. For later reference, we note that the sesquilinear form defined by polarization starting from $Q_{\alpha,S}^{(\beta)}$ w.r.t. the decompositions $\psi_1 = \phi_1 + q_1\,\chi\,G_{\lambda}$ and $\psi_2 = \phi_2 + q_2\,\chi\,G_{\lambda}$ reads
\begin{align}
Q^{(\beta)}_{\alpha,S}[\psi_1,\psi_2] 
& = Q^{(F)}_{\alpha,S}[\phi_1,\phi_2] 
- \lambda^2\, \langle \psi_1\,|\,\psi_2\rangle + \lambda^2\,\langle \phi_1\,|\,\phi_2\rangle \nonumber \\
& \quad + q_1^{*} \lf[ 2\,\braketr{\lf(\mathbf{S}\,\chi - i \nabla \chi\ri)G_{\lambda}}{\lf(-i\nabla\!+\! \mathbf{A}_{\alpha} \ri)\phi_2} + \braketl{\lf(\mathbf{S}^2\chi + \Delta \chi\ri) G_{\lambda}}{\phi_2} \ri]  \nonumber \\
& \quad + q_2\, \lf[ 2\,\braketl{\lf(-i\nabla\!+\! \mathbf{A}_{\alpha} \ri)\phi_1}{\lf(\mathbf{S}\,\chi - i \nabla \chi\ri)G_{\lambda}} + \braketr{\phi_1}{\lf(\mathbf{S}^2\chi + \Delta \chi\ri) G_{\lambda}} \ri] \nonumber \\
& \quad + q_1^{*}\,q_2\, \lf(\beta + c_{\alpha}\, \lambda^{2 \alpha} + \Xi_{\alpha,S}(\lambda) \ri)\,. \label{defQ2}
\end{align}

\begin{proof}[Proof of \cref{cor:domHb1}]
 Let us first assume $q_1 = 0$, i.e., $\psi_1 = \phi_1$, in \eqref{defQ2} above; then, upon integration by parts, the sesquilinear form reduces to
\bml{
Q^{(\beta)}_{\alpha,S}[\phi_1,\psi_2]  = \braketr{\phi_1}{H_{\alpha,S} \phi_2 + 2\,q_2\,\chi\, \mathbf{S} \!\cdot\! \lf(-i\nabla \!+\! \mathbf{A}_{\alpha}\ri)\, G_{\lambda}} \\
	+ q_2\, \lf[ \braketr{ \phi_1 }{ \lf( \mathbf{S}^2\! - \lambda^2 \ri)\,\chi\,G_{\lambda}} + 2\,\braketr{ \phi_1}{(-i \nabla \chi)\!\cdot\!\lf(-i \nabla\!+\!\mathbf{A}_{\alpha}\!+\!\mathbf{S}\ri)\,G_{\lambda}}
- \braketr{ \phi_1}{(\Delta \chi)\,G_{\lambda}} \ri] \\
	 - \lim_{r \to 0^{+}}\int_{\partial B_r}\!\! \diff \Sigma_r\;\overline{\phi_1}\,\lf[\partial_r \phi_2 + i\,\lf(\mathbf{S}\!\cdot\! \hat{\mathbf{r}}\ri)\,\phi_2 + 2\,q_2\,\lf(i\,\chi\,\mathbf{S}\!\cdot\! \hat{\mathbf{r}} + \partial_r \chi \ri)\,G_{\lambda} \ri]\,.
	\label{proeq1}
}
However, the boundary term vanishes identically: by the asymptotic relations in Eq.\,\eqref{limdomQF} of \cref{prop:QF}, we get
\bmln{
	\left|\int_{\partial B_r}\!\!d\Sigma_r\;\overline{\phi_1}\,\big[\partial_r \phi_2 + i\,(S\!\cdot\! \hat{\mathbf{r}})\,\phi_2 + 2\,q_2\,(i\,\chi\,S\!\cdot\! \hat{\mathbf{r}} + \partial_r \chi)\,G_{\lambda}\big]\right| \\
	\leqslant C  r \sqrt{\avg{|\phi_1|^2}(r)} \left[ \sqrt{\avg{\lf| \partial_r \phi_2 \ri|^2}(r)}+ \| \mathbf{S} \|_{L^{\infty}(B_r)} \sqrt{\avg{|\phi_2|^2}(r)} + 2\,|q_2|\, \lf\|\chi\, \mathbf{S} \ri\|_{\infty}\! G_{\lambda}(r)  \right] \xrightarrow[r \to 0^+]{} \;0\,.
}
On the other hand, keeping in mind the convention described in \cref{rem:L2inner}, it can be easily inferred that all the expressions within the square brackets on the second line of Eq.\,\eqref{proeq1} are finite. Thus, the condition $Q^{(\beta)}_{\alpha,S}[\phi_1,\psi_2] = \langle \phi_1\,|\,w\rangle$ for some $w : = H^{(\beta)}_{\alpha,S}\, \phi_2 \!\in\! L^2(\mathbb{R}^2)$ can be fulfilled only if \beq
	H_{\alpha,S} \phi_2 + 2\,q_2\,\chi\, \mathbf{S} \!\cdot\! \lf(-i\nabla \!+\! \mathbf{A}_{\alpha} \ri)\, G_{\lambda} \!\in\!L^2(\mathbb{R}^2)
\eeq
and 
\bml{
	w := H_{\alpha,S}\,\phi_2+ q_2 \lf[ \lf( \mathbf{S}^2\! - \lambda^2 \ri)\,\chi\,G_{\lambda}\! + 2\,\chi\, \mathbf{S} \!\cdot\!\lf(-i\nabla \!+\!\mathbf{A}_{\alpha}\ri) G_{\lambda}\! \ri.	\\
	\lf. + 2\,(-i \nabla \chi)\!\cdot\! \lf(-i \nabla\!+\!\mathbf{A}_{\alpha}\!+\!\mathbf{S} \ri) G_{\lambda}\!- (\Delta \chi) G_{\lambda} \ri]\,.\label{defw1}
}

Let now $q_1 \neq 0$. Integrating by parts and checking again that the boundary contributions vanish, we get
\bml{
Q^{(\beta)}_{\alpha,S}[\psi_1,\psi_2] = Q^{(\beta)}_{\alpha,S}[\phi_1,\psi_2] 
+ q_1^{*}\, \braketr{ G_{\lambda}}{\lf[ 2\, \lf( \mathbf{S}\,\chi + i \nabla\chi \ri)\!\cdot\! \lf(-i \nabla\!+\!\mathbf{A}_{\alpha} \ri)  + \lf( \mathbf{S}^2- \lambda^2 \ri) \chi + \Delta \chi \ri] \phi_2} \\
	+ q_1^{*}\,q_2\, \lf(\beta + c_{\alpha}\, \lambda^{2 \alpha} + \Xi_{\alpha,S}(\lambda) - \lambda^2\,\|\chi G_{\lambda}\|_{2}^{2} \ri)\,.
}
Recall the definition \eqref{defXi} of $\Xi_{\alpha,S}(\lambda)$ and that we are assuming $q_1 \neq 0$; then, demanding $Q^{(\beta)}_{\alpha,S}[\psi_1,\psi_2] = \langle \psi_1\,|\,w\rangle$ with $w$ as in Eq.\,\eqref{defw1} implies
\begin{equation}
q_2\,\lf(\beta + c_{\alpha}\, \lambda^{2 \alpha} - 2\, \braketr{ \chi\,G_{\lambda}}{ \mathbf{S} \!\cdot\!(-i \nabla)(\chi\,G_{\lambda})} \ri) = \braketr{ G_{\lambda}}{ \lf[ \lf(-i \nabla\!+\!\mathbf{A}_{\alpha} \ri)^2 + \lambda^2 \ri]\, \chi\,\phi_2} \,. \label{bcid}
\end{equation}
On one hand, exploiting the Coulomb gauge $\nabla \cdot \mathbf{S} = 0$, we obtain
\bdm
	\braketr{ \chi\,G_{\lambda}}{ \mathbf{S} \!\cdot\!(-i \nabla)(\chi\,G_{\lambda})} = -\tx\frac{i}{2}\, \disp\lim_{r \to 0^{+}} \disp\int_{\mathbb{R}^2 \setminus B_r} \diff \mathbf{x}\;\nabla \cdot \lf( \mathbf{S}\, (\chi\,G_{\lambda})^2 \ri)
= \tx\frac{i}{2}\, \disp\lim_{r \to 0^{+}} \!\int_{\partial B_r}\!\!\!\diff \Sigma_r\;( \mathbf{S} \cdot \hat{\mathbf{r}})\, (\chi\,G_{\lambda})^2 = 0\,,
\edm
where we have used the relation \eqref{Sleq} and the asymptotic expansion \eqref{asy0}, to prove that the last integral is $ \OO(r^{\nu + 1 - 2\alpha}) $, as $  r \to 0^+ $. Concerning the r.h.s. of Eq.\,\eqref{bcid}, via repeated integration by parts, we get
\bmln{
	\braketr{ G_{\lambda}}{ \lf[ \lf(-i \nabla\!+\!\mathbf{A}_{\alpha} \ri)^2 \ri]\, \chi\,\phi_2}
	= \lim_{r \to 0^{+}} \int_{\mathbb{R}^2 \setminus B_r}\!\!\!\!\diff\mathbf{x}\;G_{\lambda}\,\lf(-i\nabla \!+\! \mathbf{A}_{\alpha} \ri)^2 (\chi \phi_2) \\
	= \lim_{r \to 0^{+}}\!\int_{\mathbb{R}^2 \setminus B_r}\!\!\!\! \diff\mathbf{x}\;\overline{(-i\nabla \!+\! \mathbf{A}_{\alpha})^2 G_{\lambda}}\; \chi\, \phi_2 
+ \lim_{r \to 0^{+}}\!\int_{\partial B_r}\!\!\! \diff\Sigma_r\; \lf[G_{\lambda}\, \partial_r (\chi \phi_2) - \chi\,\phi_2\,\partial_r G_{\lambda} \ri]\,. 
}
In view of \eqref{greeeq}, the asymptotic expansion \eqref{asy0} and the fact that $\chi = 1$ in an open neighborhood of the origin, the above results entail
\bdm
	\lf(\beta + c_{\alpha}\, \lambda^{2 \alpha} \ri)\,q_2 
= \lim_{r \to 0^+} \int_{\partial B_r}\!\!\!\diff\Sigma_r\; \lf[G_{\lambda}\, \partial_r \phi_2 - \phi_2\,\partial_r G_{\lambda} \ri] = {\Gamma(\alpha) \over 2^{1 - \alpha}}\,\lim_{r \to 0^+} \left[r^{-\alpha-1}\! \int_{\partial B_r}\!\!\!\diff\Sigma_r\; \lf(r\,\partial_r \phi_2 + \alpha\, \phi_2 \ri) \right] ,
\edm
which completes the derivation of the domain \eqref{domHbe1}. 

To conclude the  proof it just remains to observe that $ \dom(\dot{H}_{\alpha,S}) \subset \dom(H_{\alpha,S}^{(\beta)}) $, since, by restricting to wave functions in \eqref{domHbe1} such that $ q = 0 $, we recover the Friedrichs extension which in turn extends $ \dot{H}_{\alpha,S} $. Hence, any operator $ H_{\alpha,S}^{(\beta)} $ identifies a self-adjoint extension of $ \dot{H}_{\alpha,S} $.

Finally, we prove that if $ \mathbf{S} \in L^{\infty}(\R^2) $ the family $ \big\{ H_{\alpha,S}^{(\beta)} \big\}_{ \beta \in \mathbb{R} } $ exhausts all self-adjoint extensions of $ \dot{H}_{\alpha,S} $ in $ L_{\mathrm{even}}^2(\R^2) $. To this purpose we first notice that \eqref{HalSC} implies that we can write
$$
\dot{H}_{\alpha,S} = \dot{H}_{\alpha,0} + 2\,\mathbf{S} \cdot \lf( -i \nabla + \mathbf{A}_{\alpha} \ri) + \mathbf{S}^2\,.
$$
For any $\psi \in C^{\infty}_{c}(\mathbb{R}^2 \setminus \{\mathbf{0}\})$ and for all $\varepsilon > 0$, using the Cauchy-Schwarz inequality together with the basic inequality $\sqrt{u\, v} \leq \varepsilon\,u + {1 \over 2\varepsilon}\,v$ ($u,v > 0$), we get
\beq
	\|(-i \nabla + \mathbf{A}_{\alpha}) \psi \|_{2} = \sqrt{ \braketr{ \psi }{ \dot{H}_{\alpha,0}\, \psi } } \leq \varepsilon\, \big\|\dot{H}_{\alpha,0} \psi\big\|_{2} + {1 \over 2 \varepsilon}\,\|\psi\|_{2}\,,
\eeq
which, recalling that $C^{\infty}_{c}(\mathbb{R}^2 \setminus \{\mathbf{0}\})$ is dense in $\dom\big(\dot{H}_{\alpha,0}\big)$, allows us to infer, for all $\psi \in \dom\big(\dot{H}_{\alpha,0}\big)$,
\bml{
	\lf\|\lf(2\,\mathbf{S} \cdot \lf( -i \nabla + \mathbf{A}_{\alpha} \ri) + \mathbf{S}^2\ri) \psi\ri\|_{2} \leq \|\mathbf{S}\|_{\infty}\, \|(-i \nabla + \mathbf{A}_{\alpha}) \psi\|_{2} + \|\mathbf{S}\|_{\infty}^2 \|\psi\|_{2} \\
	\leq \, \varepsilon\,\|\mathbf{S}\|_{\infty}\, \big\|\dot{H}_{\alpha,0} \psi\big\|_{2} + \|\mathbf{S}\|_{\infty}\! \lf(\tx{1 \over 2 \varepsilon} + \|\mathbf{S}\|_{\infty}\!\ri) \|\psi\|_{2}\,.
}
Fixing arbitrarily $\varepsilon < 1/\|\mathbf{S}\|_{\infty}$, the latter chain of inequalities shows that $2\,\mathbf{S} \cdot \lf( -i \nabla + \mathbf{A}_{\alpha} \ri) + \mathbf{S}^2$ is $\dot{H}_{\alpha,0}$\,-bounded with relative bound smaller than $1$. 
Then, by a variant of Kato-Rellich theorem (see, {\it e.g.}, \cite[Thm. 9, p. 100]{BiSo} and \cite[Eq.(1.1) and related references]{Kis}) it follows that the deficiency indices of $ \dot{H}_{\alpha,S} $ and $ \dot{H}_{\alpha,0} $ coincide. Since $ \dot{H}_{\alpha,0} \upharpoonright L_{\mathrm{even}}^2(\R^2) $ has deficiency indices $(1,1)$ (see \cite{AdTe,CoOd}), the previous arguments suffice to deduce the thesis.
\end{proof}

\subsection{Singular perturbations ($ \mathbf{S} $ discontinuous)}

As in the previous section, let us first verify that the quadratic form is well-posed, with the same convention as in \cref{rem:L2inner}. Preliminarily, we observe that, combining \eqref{eq:defzetan} with the asymptotics of $ G_{\la} $ at the origin given in \eqref{asy0}, we obtain
$$
\chi\,\zeta_{\alpha}G_{\lambda} = \left\{\!\begin{array}{ll}
\displaystyle{{\Gamma(\alpha) \over 2^{1 - \alpha}}\, r^{-\alpha} - {\Gamma(1-\alpha)\,\lambda^{2 \alpha} \over 2^{1 + \alpha}\,\alpha}\,r^{\alpha} + \mathcal{O}(r^{2-\alpha})\,,} &	\displaystyle{\mbox{for\, $0 < \alpha < 1/2$}\,,}\vspace{0.1cm}\\
\displaystyle{\sqrt{{\pi \over 2}}\, r^{-1/2} 
+ \sqrt{{\pi \over 2}}\,r^{1/2} \Big(S_{\perp}(\mathbf{0})\, \log r - \lambda \Big)
+ \mathcal{O}(r^{3/2}\log r)
\,,} &	\displaystyle{\mbox{for\, $\alpha = 1/2$}\,,}\vspace{0.1cm}\\
\displaystyle{
{\Gamma(\alpha) \over 2^{1 - \alpha}}\, r^{-\alpha} 
- {\Gamma(1+\alpha) \over 2^{1 - \alpha}(\alpha -1/2)}\, S_{\perp}(\mathbf{0})\, r^{1-\alpha}
- {\Gamma(1-\alpha)\,\lambda^{2 \alpha} \over 2^{1 + \alpha}\,\alpha}\,r^{\alpha} 
+ \mathcal{O}(r^{2-\alpha})
\,,}	&	\displaystyle{\mbox{for\, $1/2 < \alpha < 1$}\,.}
\end{array}\right.
$$
Therefore, the main modification to the defect function is related to the asymptotic behavior of the next-to-leading order term, which, for $ \alpha \geq 1/2 $, behaves as $ r^{1/2} \log r $ or $ r^{1-\alpha} $ instead of $ r^{\alpha} $, respectively.
	
Consider now the expression $\braketr{ \phi_{\lambda}}{\chi\,\Delta \zeta_{\alpha}\, G_{\lambda}} $ in the third line of \eqref{Qbeta2}. For $0 < \alpha < 1/2$ we have $\Delta \zeta_{\alpha} = 0$, so the said term vanishes identically. Contrarily, for $1/2 \leqslant \alpha < 1$ it appears that $ \Delta \zeta_{\alpha} $ is singular at the origin and $\braketr{ \phi_{\lambda}}{\,\chi\,\Delta \zeta_{\alpha}\, G_{\lambda}} $ must be understood as a pairing between the weighted space $L^2(\mathbf{R}^2,|\mathbf{x}|^{-2} d\mathbf{x})$ and its dual $L^2(\mathbf{R}^2,|\mathbf{x}|^{2} d\mathbf{x})$. To account for this claim, it suffices to notice that $\chi\,(|\mathbf{x}|\, \Delta \zeta_{\alpha})\,G_{\lambda} \!\in\! L^{2}(\mathbb{R}^2)$ and to recall that $\mathbf{A}_{\alpha} \phi_{\lambda} \!\in\! L^2(\mathbb{R}^2)$ for any $\phi_{\lambda} \!\in\! \mbox{dom}\big(Q_{\alpha,S}^{(F)}\big)$ (see \eqref{domQF}), so that
\bmln{
	\lf| \braketr{\phi_{\lambda}}{\,\chi\,\Delta \zeta_{\alpha}\,G_{\lambda}} \ri| \leqslant \lf\|\phi_{\lambda} \ri\|_{L^2\left(\mathbf{R}^2,\,|\mathbf{x}|^{-2} d\mathbf{x}\right)}\, \lf\|\chi\, \Delta \zeta_{\alpha}\, G_{\lambda} \ri\|_{L^2(\mathbf{R}^2,\,|\mathbf{x}|^2 d\mathbf{x})} \\
	= \left(\int_{\mathbb{R}^2}\!\!\diff\mathbf{x}\; |\mathbf{x}|^{-2}\, |\phi_{\lambda}|^2 \right)^{\!\!1/2}\! \left(\int_{\mathbb{R}^2}\!\!\diff\mathbf{x}\; \chi^2\,(|\mathbf{x}|\,\Delta \zeta_{\alpha})^2\, G_{\lambda}^2\right)^{\!\!1/2}\!
= \tx{1\over\alpha}\; \lf\|\mathbf{A}_{\alpha}\phi_{\lambda} \ri\|_{2}\;\lf\|\,\chi\,(|\mathbf{x}|\,\Delta \zeta_{\alpha})\,G_{\lambda} \ri\|_{2}\,.
}

On the other hand, let us examine the potentially troublesome term in the second line of \eqref{defXi2}. Using \eqref{SAlim} together with the simple identities
\begin{eqnarray}
	\nabla\zeta_{\alpha} = \left\{\!\begin{array}{ll}
\displaystyle{0\,,} &	\displaystyle{\mbox{for\, $0 < \alpha < 1/2$}\,,}\vspace{0.1cm}\\
\displaystyle{S_{\perp}(\mathbf{0})\,(\log r + 1)\,\hat{\mathbf{x}}\,,} &	\displaystyle{\mbox{for\, $\alpha = 1/2$}\,,}\vspace{0.1cm}\\
\displaystyle{- \,{\alpha\, S_{\perp}(\mathbf{0}) \over \alpha -1/2}\,\hat{\mathbf{x}}\,,}	&	\displaystyle{\mbox{for\, $1/2 < \alpha < 1$}\,,}
\end{array}\right. \label{eq:gradzeta}	\\
	\Delta \zeta_{\alpha}(r) = \left\{\!\begin{array}{ll}
\displaystyle{0\,,} &	\displaystyle{\mbox{for\, $0 < \alpha < 1/2$}\,,}\vspace{0.1cm}\\
\displaystyle{S_{\perp}(\mathbf{0})\,{\log r + 2 \over r}\,,} &	\displaystyle{\mbox{for\, $\alpha = 1/2$}\,,}\vspace{0.1cm}\\
\displaystyle{- \,{\alpha\, S_{\perp}(\mathbf{0}) \over \alpha -1/2}\,{1 \over r}\,,}	&	\displaystyle{\mbox{for\, $1/2 < \alpha < 1$}\,,}
\end{array}\right. \label{eq:lapzeta}
\end{eqnarray} 
following from \eqref{eq:defzetan}, we get, for any $0 < \alpha <1$,
\begin{equation}
2\,\mathbf{S}\!\cdot\!\mathbf{A}_{\alpha}\, \zeta_{\alpha}^2\, G_{\lambda} - \lf(\zeta_{\alpha}\, \Delta \zeta_{\alpha}\ri) G_{\lambda} - 2\,\zeta_{\alpha}\,\nabla \zeta_{\alpha} \cdot\! \nabla G_{\lambda}
= \mathcal{O}(r^{\nu-1-\alpha})\, ,
\end{equation}
where $ \nu > 2 \alpha - 1 $, if $ \alpha \in [1/2,1) $, and $\nu = 0$ otherwise (recall \eqref{Shyp 2a}). In view of the above asymptotics, the expression in the second line of \eqref{defXi2} can be given a meaning as a duality pairing between  $L^2(\mathbb{R}^2,\,|\mathbf{x}|^{\,-\,(1-\nu)}\diff\mathbf{x})$ and $L^2(\mathbb{R}^2,\,|\mathbf{x}|^{1-\nu} \diff\mathbf{x})$.

\begin{proof}[Proof of \cref{prop:Qbeta2}]	Once the expression of the quadratic form is shown to be well-posed, the proof of the result follows the very same lines of the proof of \cref{prop:Qbeta}. We omit the details for the sake of brevity.
\end{proof}

\begin{proof}[Proof of \cref{cor:domHb12}] We only highlight here the major changes to the argument in the proof of \cref{cor:domHb1}. In particular, in the derivation of the operator domain, one must replace $\chi$ with $\chi\,\zeta_{\alpha}$. Hence, one has to verify that the term $\braketr{ \chi\, \zeta_{\alpha}\,G_{\lambda}}{ \mathbf{S} \!\cdot\!(-i \nabla)(\chi\, \zeta_{\alpha}\,G_{\lambda})}$ does not give any boundary contribution:
\bml{
	\braketr{ \chi\, \zeta_{\alpha}\,G_{\lambda}}{ \mathbf{S} \!\cdot\!(-i \nabla)(\chi\, \zeta_{\alpha}\,G_{\lambda})} = -\tx\frac{i}{2}\, \disp\lim_{r \to 0^{+}} \disp\int_{\mathbb{R}^2 \setminus B_r}\!\!\! \diff \mathbf{x}\;\nabla \cdot \lf( \mathbf{S} (\chi\, \zeta_{\alpha}\,G_{\lambda})^2 \ri) \\
= \tx\frac{i}{2}\, \disp\lim_{r \to 0^{+}} \!\int_{\partial B_r}\!\!\!\diff \Sigma_r\;( \mathbf{S} \cdot \hat{\mathbf{r}})\, (\chi\, \zeta_{\alpha}\,G_{\lambda})^2
= i\,\pi\, \disp\lim_{r \to 0^{+}}  \left(r\, G_{\lambda}^2(r) \!\fint_{\partial B_r}\!\!\!\diff \Sigma_r\;\mathbf{S} \cdot \hat{\mathbf{r}} \right)	\\
= {i\,\pi\,\Gamma^2(\alpha) \over 2^{2-2\alpha}}\, \disp\lim_{r \to 0^{+}} \left[r^{1-2\alpha}\,  \avg{S_{\parallel}}\!(r) \right] = 0,
}
thanks to \eqref{Shyp 2a}.
Taking this into account and noting as well that the expression $\partial_{r} (\chi\,\zeta_{\alpha})$ in general does not vanish for $r \to 0^{+}$, in this case \eqref{bcid} implies
\bml{
	\lf(\beta + c_{\alpha}\, \lambda^{2 \alpha} \ri) q_2 
= \lim_{r \to 0^+} \int_{\partial B_r}\!\!\!\diff\Sigma_r\, \lf[\partial_r(\chi\,\zeta_{\alpha}\, \phi_{2})\,G_{\lambda} - \chi\,\zeta_{\alpha}\, \phi_{2}\,\partial_r G_{\lambda} + i\,q_2\,(\mathbf{S}\cdot \hat{\mathbf{r}})\,\chi^2 \zeta_{\alpha}^2 G_{\lambda}^2\ri] \\
= \lim_{r \to 0^+} \left[2 \pi r \left((\partial_r \zeta_{\alpha}) \,G_{\lambda} \avg{\phi_{2}} + \zeta_{\alpha}\,G_{\lambda} \avg{\partial_r \phi_{2}} - \zeta_{\alpha}\,\partial_r G_{\lambda} \avg{\phi_{2}}  \right)\right] .
}
Then, the identity in the third line of \eqref{domHbe12} follows recalling the asymptotics \eqref{asy0} and noting that the basic assumption \eqref{Shyp} grants $|\avg{\mathbf{S} \cdot \hat{\mathbf{r}}}(r)| \leqslant \|\mathbf{S}\|_{L^{\infty}(B_r)} = \mathcal{O}(1)$, while the relations in \eqref{limdomQF} and the Cauchy-Schwarz inequality imply $|\avg{\phi_2}\!(r)| \leqslant \sqrt{\avg{|\phi_2|}\!(r)} \to 0$ and $r\,|\avg{\partial_{r}\phi_2}\!(r)| \leqslant \sqrt{r^2 \avg{|\partial_{r}\phi_2|}\!(r)} \to 0$.
\end{proof}

\subsection{Von Neumann theory for special $ \mathbf{S} $}
\label{sec: special proof} 		

We prove here the classification in \cref{prop: special}. We remark that, by \eqref{Shyp 3}, Lipschitz continuity of $ S $ yields
\beq
	\label{eq: lipschitz}
	\big|S(r) - S(0)\big| \leqslant C\,r\qquad \mbox{for some\, $C > 0$\, and for all\, $r \in [0,r_0)$}\,.
\eeq

\begin{proof}[Proof of \cref{prop: special}]
As anticipated, the key idea is to decompose in cylindrical harmonics the operator to get
\begin{equation}
	H_{\alpha,S} = \sum_{k \in \mathbb{Z}} V^{-1}\, h_{\alpha,S}^{(k)}\, V \otimes \mathbf{1}\;, \label{Hsph}
\end{equation}
where the radial differential operators are given by
\begin{equation}
	h_{\alpha,S}^{(k)} := -\,{d^2\, \over dr^2} + {\big(k + \alpha + r\,S(r)\big)^2 - 1/4 \over r^2}\;. \label{hkrad}
\end{equation}
It is convenient to consider the following decomposition:
\beq
	h_{\alpha,S}^{(k)} = L_{\alpha,S}^{(k)} + W_{\alpha,S}^{(k)}(r), \qquad k \in \mathbb{Z}\;; \label{hkrad2}
\eeq
where the operators $ L_{\alpha,S}^{(k)} $ are defined in \eqref{eq: Lalpha} and $ W_{\alpha,S}^{(k)} $ are the radial potentials
\beq
	 W_{\alpha,S}^{(k)}(r) := {2(k + \alpha)\,\big(S(r) - S(0)\big) \over r} + S^2(r)\,. 
\eeq

By \eqref{Shyp 3} and \eqref{eq: lipschitz}, for any $ k\in \Z $, $ W_{\alpha,S}^{(k)} \in L^{\infty}(\R_+) $, which allows to treat it as a bounded and thus infinitesimally Kato-small perturbation of $ L_{\alpha,S}^{(k)}  $, while the discussion of the latter operators was already done in \cite{DR18,DR20}. Let us then denote with $\dot{h}_{\alpha,S}^{(k)}$ the closure of the densely-defined, symmetric operator $h_{\alpha,S}^{(k)}\!\upharpoonright\!C^{\infty}_{c}(\mathbb{R}_{+})$, and define in the same way the corresponding analogue $\dot{L}_{\alpha,S}^{(k)}$. For all $k \in \mathbb{Z}\setminus \{-1,0\}$, $\dot{L}_{\alpha,S}^{(k)}$ is self-adjoint on $L^2(\mathbb{R}_{+},dr)$ \cite[Prop. 2.1, item (v)]{DR20}. Since $W_{\alpha,S}^{(k)}$ is a bounded perturbation, the same can be said for the associated operator $\dot{h}_{\alpha,S}^{(k)}$ by Kato-Rellich theorem. More precisely, $\dot{h}_{\alpha,S}^{(k)}$ coincides with the Friedrichs extension $h_{\alpha,S}^{(F,k)}$, whose domain is
\begin{equation}
 	\dom\big(h_{\alpha,S}^{(k,F)}\big) = \big\{\varphi \in L^2(\mathbb{R}_{+},dr) \,\big|\, h_{\alpha,S}^{(k)}\varphi\in L^2(\mathbb{R}_{+},dr)\big\}\,.
\end{equation}
Incidentally, let us also mention that, by \cite[Prop. 2.1]{DR20}, 
\begin{equation*}
	\dom\big(h_{\alpha,S}^{(k,F)}\big) \!\subset\! \lf\{ \varphi \!\in\! C^{1}(\mathbb{R}_{+})\,\Big|\, \varphi(r) = o(r^{3/2})\,, \varphi'(r) = o(r^{1/2}) \mbox{ as } r \to 0^{+},\; \varphi(r) \xrightarrow[r \to +\infty]{} 0 \ri\}\,.
\end{equation*}

For either $k = 0$ or $k = -1$, $\dot{L}_{\alpha,S}^{(k)}$ has deficiency indices $(1,1)$ (see, e.g., \cite{Bulla} or \cite[Prop. 3.1, item (v)]{DR18}). It can be checked by direct inspection that the deficiency functions $g_{\alpha,\pm}^{(k)}$ fulfilling
\begin{equation}
\big(\dot{L}_{\alpha,S}^{(k)}\big)^{*}\, g_{\alpha,\pm}^{(k)} = \pm i\,g_{\alpha,\pm}^{(k)}\,,\label{defi}
\end{equation}
are given by \eqref{eq: galpha}. The functions $g_{\alpha,\pm}^{(k)}$ are unique up to a constant factor.
To account for the above statements note that all the distributional solutions of Eq.\,\eqref{defi} are of the form
\begin{equation*}
g_{\alpha,\pm}^{(k)}(r) = C_1\, M_{-\, e^{\pm i \pi/4} (\alpha + k) S(0),\, |\alpha + k|}\big(2\,e^{\mp i \pi/4}\, r\big) + C_2\, W_{-\, e^{\pm i \pi/4} (\alpha + k) S(0),\, |\alpha + k|}\big(2\,e^{\mp i \pi/4}\, r\big)\,,
\end{equation*}
where $M_{\kappa,\nu}(z)$ and $W_{\kappa,\nu}(z)$ are the Whittaker functions\footnote{Introducing the complex variable $z := 2 \sqrt{\mp i}\; r$ (with the determination $\sqrt{\mp i} = e^{\mp i \pi/4}$, yielding $\Re z > 0$) and setting $\widetilde{g}_{\alpha,\pm}^{(k)}(z) := g_{\alpha,\pm}^{(k)}(z/2\sqrt{\mp i})$, the defect equation \eqref{defi} can be recast in the standard Whittaker form \cite[Eq.\,13.14.1]{NIST}
	\[ 
	\left(-\,{d^2\, \over dz^2} + {(k + \alpha)^2 - 1/4 \over z^2} + {e^{\pm i \pi/4} (k + \alpha)S(0) \over z} + {1 \over 4}\right) \widetilde{g}_{\alpha,\pm}^{(k)}(z) = 0\,. 
	\]} defined in \cite[\S\,13.14]{NIST}.
These special functions are in fact equivalent, up to rescaling, to the special functions denoted by $ \mathcal{I}_{\beta,m} $ and $ \mathcal{K}_{\beta,m} $ in \cite{DR18}.
Square-integrability near the origin is always granted, since \cite[\S 13.14(iii)]{NIST}
\begin{equation}\label{asygpm0}
g_{\alpha,\pm}^{(0)}(r) \underset{r \to 0^+}{\sim} r^{1/2-\alpha} \,, \qquad 
g_{\alpha,\pm}^{(-1)}(r) \underset{r \to 0^+}{\sim} r^{\alpha-1/2}.
\end{equation}
On the other hand, from \cite[Eqs.\,13.14.20]{NIST} we infer that (notice that $\Re\big(e^{\pm i \pi/4}\big) = 1/\sqrt{2} > 0$)
\[
	M_{-\, e^{\pm i \pi/4} (\alpha + k) S(0),\, |\alpha + k|}\big(2\,e^{\mp i \pi/4}\, r\big) \underset{r \to +\infty}{\sim} e^{e^{\mp i \pi/4}\, r}\; \big(2\,e^{\mp i \pi/4}\, r\big)^{e^{\pm i \pi/4} (\alpha + k) S(0)}\,.
\]
Thus, to ensure square-integrability at infinity of the solutions $g_{\alpha,\pm}^{(k)}$, we must set $C_1 = 0$.

Now consider the relation \eqref{Hsph} and recall that $\sum_{k \in \mathbb{Z}} V^{-1}\, W_{\alpha,S}^{(k)}\, V \otimes \mathbf{1}$ is a bounded perturbation of $\sum_{k \in \mathbb{Z}} V^{-1}\, L_{\alpha,S}^{(k)}\, V \otimes \mathbf{1}$. Then, from the above results and a variant of Kato-Rellich theorem \cite[p.100, Ch.4, Thm.9]{BiSo}, \cite[Eq.(1.1) and related references]{Kis} it follows that $\dot{H}_{\alpha,S}$ has deficiency indices $(2,2)$, and the equation $\dot{H}_{\alpha,S}\,\Upsilon = \pm i\,\Upsilon$ has two independent solutions given by \eqref{Upsilondef} forming a basis for the two-dimensional deficiency subspaces $ \ker(\dot{H}_{\alpha,S}^{*} \mp i)$. Applying
standard Von Neumann theory \cite{ReSi1}, we get the result.
\end{proof}

\bigskip

\noindent
{\bf Data Availability.} Data sharing is not applicable to this article as no new data were created or analyzed in this study.


\end{document}